\documentclass[showpacs, aps, prb, unsortedaddress,showkeys,longbibliography,comma,authoryear,round]{revtex4-1}

\usepackage{amssymb}
\usepackage{latexsym}
\usepackage{amsmath}
\usepackage{graphicx}
\usepackage{float}
\usepackage{xcolor}
\usepackage{amsthm}
\setcitestyle{authoryear,open={(},close={)}}

\newtheorem{theorem}{Theorem}[section]

\newtheoremstyle{named}{}{}{\itshape}{}{\bfseries}{.}{.5em}{\thmnote{#3's }#1}
\theoremstyle{named}
\newtheorem*{namedtheorem}{Theorem}

\begin{document}
\newpage
\title{Causality and Passivity: from Electromagnetism and Network Theory to Metamaterials}

\date{\today}
\author{Ankit Srivastava}
\thanks{Corresponding Author}
\affiliation{Department of Mechanical, Materials, and Aerospace Engineering,
Illinois Institute of Technology, Chicago, IL, 60616
USA}
\email{asriva13@iit.edu}

\begin{abstract}
In this review, we take an extensive look at the role that the principles of causality and passivity have played in various areas of physics and engineering, including in the modern field of metamaterials. The aim is not to provide a comprehensive list of references as that number would be in the thousands, but to review the major results and contributions which have animated these areas and to provide a unified framework which could be useful in understanding the developments in different fields. Towards these goals, we chart the early history of the field through its dual beginnings in the analysis of the Sellmeier equation and in Hilbert transforms, giving rise to the far reaching dispersion relations in the early works of Sokhotskii, Plemelj, Kramers, Kronig, and Titchmarsh. However, these early relations constitute a limited result as they only apply to a restricted class of transfer functions. To understand how this restriction can be lifted, we take a quick detour into the distributional analysis of Schwartz, and discuss the dispersion relations in the context of distribution theory. This approach expands the reach of the dispersion analysis to distributional transfer functions and also to those functions which exhibit polynomial growth properties. To generalize the results even further to tensorial transfer functions, we consider the concept of passivity - originally studied in the theory of electrical networks. We clarify why passivity implies causality and present generalized dispersion relations applicable to transfer functions which are distributional, tensorial, and possibly exhibiting polynomial growth. Subsequently, as special cases, we present examples of dispersion relations from several areas of physics including electromagnetism, acoustics, seismology, reflectance measurements, and scattering theory. We discuss sum rules which follow from the infinite integral dispersion relations and also how these integrals may be simplified either by truncating them under appropriate assumptions or by replacing them with derivative relations. These derivative relations, termed derivative analyticity relations, form the basis of the so called nearly-local approximations of the dispersion relations which are extensively employed in many fields including acoustics. Finally, we review the clever applications of ideas from causality and passivity to the recent field of metamaterials. In many ways, these ideas have provided limits to what can be achieved in metamaterial property design and metamaterial device performance.

\end{abstract}
\keywords{causality,passivity,metamaterials}

\maketitle

\section{Introduction}

If a cause-effect relation adopts a convolution form, then the assumption that the effect cannot exist before its cause -- the colloquial statement of causality -- has strong implications for the transfer function of the relationship. Such transfer functions are ubiquitous in physics and engineering. In appropriate limits, they are the dielectric permittivity and magnetic permeability of electromagnetic materials, the density and bulk modulus of acoustic media, the impedance and admittance of electrical circuits, and the compliance and stiffness of solid materials - just to name a few. The causality restrictions apply to all of them equally and ensure that the real and imaginary parts of their Fourier transforms are not independent quantities but are derivable from each other. This interdependence is the crux of the famous Kramers-Kronig dispersion relations which connect the real and imaginary parts of the Fourier transform of causal transfer functions to each other through a Hilbert transform.

In this review, we consider this idea of causality from both historical and modern perspectives. Even though causality lends itself quickly to rather complex mathematics, it started out with intuitive ideas and physical examples. In section \ref{sec:introduction}, we motivate the paper with the simple example of a single degree of freedom forced damped oscillator. The solution for the response of this problem clarifies all the essential features of the problem of causality. In this problem, the presence of damping ensures that the poles of the transfer function are in the lower half of the complex frequency plane and, therefore, that the system is causal. The location of the poles in the lower half immediately results in the emergence of the Kramers-Kronig dispersion relations. The equivalence of passivity, causality, and dispersion relation is, therefore, evident from this example. As we describe later on, this simple model -- termed the Lorentz oscillator model -- is important to the early development of the field of metamaterials. In section \ref{subsec:point}, we describe the necessary restrictions on a linear and causal transfer function for it to satisfy the classical Kramers-Kronig dispersion relations. The main result in this section is the Titchmarsh's theorem. The Titchmarsh's theorem only applies to point functions with restricted growth properties. To expand its reach, we consider causality applied to generalized functions or distributions in section \ref{subsec:distributions}. The main result here is an expression for the generalized Hilbert transform (Eq. \ref{eKramersKronigDist}), which is the most general form of the dispersion relations possible for scalar valued transfer functions. We conclude this section with arguments which show why passivity and causality are equivalent (\ref{subsec:passcaus}).

Transfer functions in many areas of physics are not scalar valued. For example, the stiffness tensor of solids is a fourth order tensor and even the {mass density (in the context of metamaterials)} is a second order tensor. The arguments of causality and passivity should, therefore, be extended to tensorial transfer functions. We address this in section \ref{sec:scattImm} with a description of some early results from network theory where the concern was passive networks characterized by tensorial transfer functions. It is in this section that the two equivalent definitions of passivity are described: the scattering and immittance formalisms. While this equivalence is presented in the context of network theory and it may appear like a mathematical trick, it has fundamental implications in other areas of physics as well. In general, the immittance formalism is connected to causality requirements on material properties whereas the scattering formalism is connected to causality requirements on the scattering of waves. However, macroscale material properties are nothing but homogenized descriptions of complex scattering phenomenon at the micro-scale. Therefore, the causality requirements at the two scales are connected to each other, which is also manifested in the equivalence between the immittance and scattering formalisms. Section \ref{sec:Passive} completes this discussion with theorems \ref{tCPScatt},\ref{tCPImm},\ref{tHerglotz} which clarify the connections between passive tensorial transfer functions, Herglotz functions, and appropriate dispersion relations applicable to them.

In section \ref{sec:dispersion}, we take a deeper look at causality mandated dispersion relations which, in their most general form, are given in Eqs. (\ref{eHilbertGenDisSub},\ref{eKramersKronigDistMat},\ref{eKramersKronigDistMatPositiveF}). Specializations of these relations are used in numerous areas of physics, some examples of which are given in section \ref{subsec:examples}. These examples are taken from electromagnetism, acoustics, seismology, scattering of waves, and reflectance measurements. Since the dispersion relations are infinite integral expressions, there is a general interest in trying to simplify them through various techniques. Sometimes these techniques result in useful sum-rules which, for example, allow us to estimate the amount interstellar dust in space, and sometimes they result in derivative relations and nearly local approximations of the dispersion relations. These techniques are reviewed in section \ref{subsec:approx}.

In the final section (\ref{sec:metamaterials}), we review the applications of causality and passivity in metamaterials research. In doing so, we summarize the answers to some important questions in the field. Are metamaterial properties causally consistent? Can negative material properties be achieved without losing significant amount of energy in dissipation? What are the constraints on achievable metamaterial properties coming from passivity and causality? Do invisibility cloaks really scatter less than the uncloaked object, and is it even possible to design a perfect cloak?

\section{Causality and Passivity}\label{sec:introduction}
\subsection{Motivating example}
\begin{figure}[htp]
\centering
\includegraphics[scale=.9]{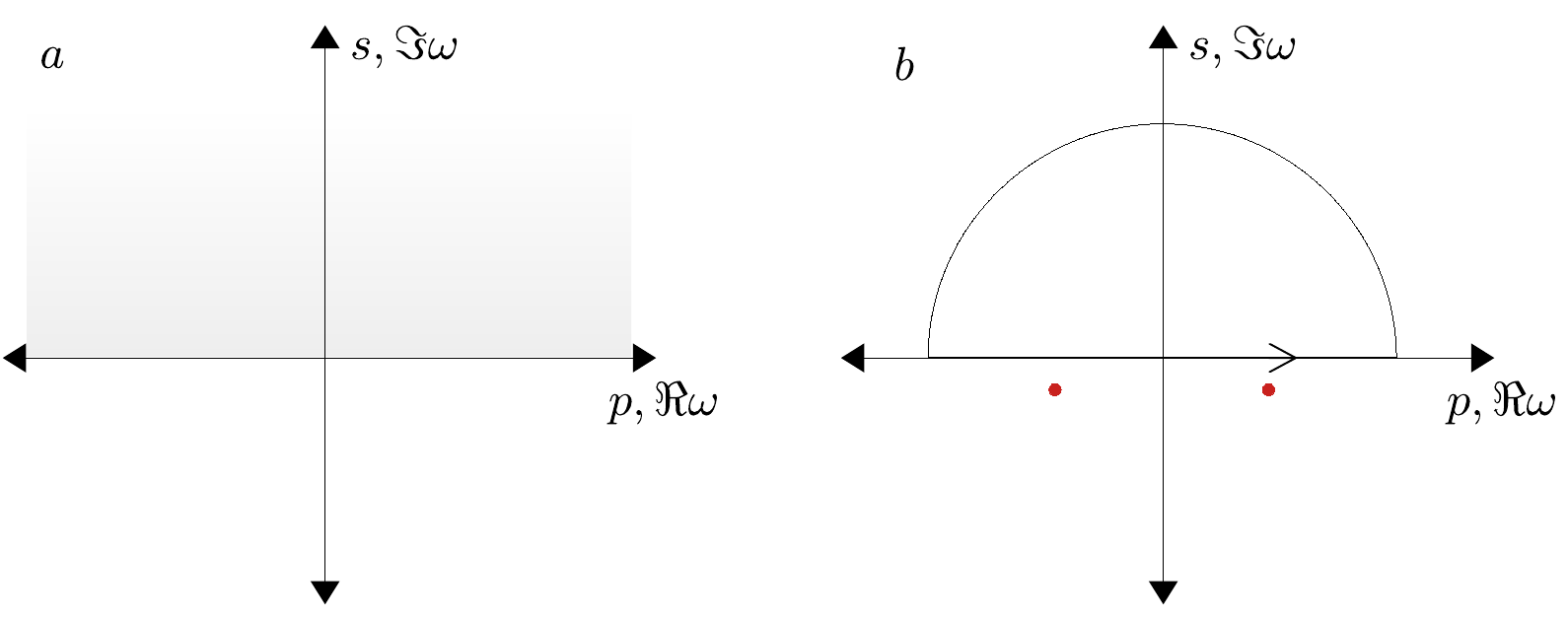}
\caption{a. Schematic of the complex frequency plane used in this paper; upper half plane $\Im\omega^+$ is represented by the shaded region and corresponds to $s,\Im\omega>0$, b. Contour used for integration in the single degree of freedom problem with the red dots representing the poles of the problem.}\label{fhalfPlane}
\end{figure}

Consider the single degree of freedom forced differential equation:
\begin{eqnarray}
\label{eSDOF}
\ddot{x}+\gamma \dot{x}+\omega_0^2x=f(t)
\end{eqnarray}
whose solution is given by the Duhamel's integral:
\begin{eqnarray}
x(t)=\int_{-\infty}^{\infty}g(t-\tau)f(\tau)\mathrm{d}\tau\equiv g(t)*f(t)
\end{eqnarray}
where * is the convolution operator and $g(t)$ is the Green's function of the problem. Taking the Fourier transform of the above, we have:
\begin{eqnarray}
X(\omega)=G(\omega)F(\omega)
\end{eqnarray}
where $X,G,F$ are the Fourier transforms of $x,g,f$, respectively. The Fourier transform pair is given by:
\begin{eqnarray}
\label{eFourierPoint}
\nonumber X(\omega)=\int_{-\infty}^\infty x(t)e^{i\omega t}dt\\
x(t)=\frac{1}{2\pi}\int_{-\infty}^\infty X(\omega)e^{-i\omega t}dt
\end{eqnarray}
In general, the frequency $\omega$ will be considered as complex with real ($p=\Re\omega$) and imaginary ($s=\Im\omega$) parts such that $\omega=p+is$ (Fig. \ref{fhalfPlane}a). We have:
\begin{eqnarray}
\label{eLorentzOscillator}
G(\omega)=\frac{1}{\omega_0^2-\omega^2-i\omega\gamma}
\end{eqnarray}
which is also called a Lorentz model. 
$\gamma$ represents the dissipation in the system and if the system is passive ($\gamma>0$), then both the poles of $G(\omega)$ lie in the plane $\Im\omega<0$ (lower half plane denoted by $\Im\omega^-$; see Fig. \ref{fhalfPlane}). Now, $g(t)$ is the inverse Fourier transform of $G(\omega)$:
\begin{eqnarray}
\label{inverseFourierPoint}
g(t)=\frac{1}{2\pi}\int_{-\infty}^{\infty}G(\omega)e^{-i\omega t}d\omega
\end{eqnarray}
If $t<0$, then $e^{-i\omega t}$ is bounded in the upper half plane (denoted by $\Im\omega^+$), and the above integral may be evaluated by considering a contour which extends in $\Im\omega^+$. Since there are no poles of $G(\omega)$ in $\Im\omega^+$, the above integral will evaluate to zero. Since $g(t)=0,t<0$ we have:
\begin{eqnarray}
x(t)=\int_{-\infty}^{t}g(t-\tau)f(\tau)\mathrm{d}\tau,
\end{eqnarray}
showing that the value of $x$ at any time $t$ can depend only upon the values of $f$ at previous time instances. Additionally, the Fourier transform of the response function, $G(\omega)$, is analytic in $\Im\omega^+$ which allows us to derive important expressions for $G(\omega)$. Consider Cauchy's integral theorem: 
$$
G(\omega)=\frac{1}{2\pi i}\int_\Gamma \frac{G(\omega')}{\omega'-\omega}d\omega'
$$
where $\Gamma$ is a closed contour encircling $\omega$. This closed contour can be chosen to include the real $\omega$ line and extended into $\Im\omega^+$. Under certain restrictions on $G(\omega)$ (to be discussed later), this process results in the following:
$$
G(\omega)=\frac{1}{2\pi i}\int_{-\infty}^{\infty} \frac{G(\omega')}{\omega'-\omega}d\omega';\quad \Im{\omega}>0
$$
The above is a direct consequence of the fact that $G(\omega)$ is analytic in $\Im\omega^+$. In other words, the value of $G(\omega)$ is expressed in terms of its values over the real line. The above can be evaluated when $\omega$ is on the real line. In this case, there is a singularity of the integrand at $\omega'=\omega$, and the integral equals:
$$
G(\omega)=\frac{1}{\pi i}\mathcal{P}\int_{-\infty}^{\infty} \frac{G(\omega')}{\omega'-\omega}d\omega';\quad \Im{\omega}=0
$$
where $\mathcal{P}$ denotes the Cauchy Principal Value. Separating the real and imaginary parts of $G(\omega)$, we have (for $\Im\omega=0$):
\begin{eqnarray}
\label{eKKSDOF}
\nonumber \Re G(\omega)=\frac{1}{\pi}\mathcal{P}\int_{-\infty}^{\infty} \frac{\Im G(\omega')}{\omega'-\omega}d\omega'\\
\Im G(\omega)=-\frac{1}{\pi}\mathcal{P}\int_{-\infty}^{\infty} \frac{\Re G(\omega')}{\omega'-\omega}d\omega'
\end{eqnarray}
Therefore, in the narrow problem described by Eq. (\ref{eSDOF}), assumption of passivity has very important implications. First, it ensures that the time domain Green's function vanishes for $t<0$, which means that $x(t)$ can depend only on the previous values of $f(t)$ (statement of causality). Second, it implies that the Fourier transform of the Green's function is analytic in $\Im\omega^+$. Finally, with appropriate restrictions on $G(\omega)$, it also means that the real and imaginary parts of $G(\omega)$ are connected to each other using Eqs. (\ref{eKKSDOF}). It is well known that if we take the real and imaginary parts of the Lorentz model (\ref{eLorentzOscillator}):
\begin{eqnarray}
\label{eLorentzRealImag}
\Re G(\omega)=\frac{\omega_0^2-\omega^2}{(\omega_0^2-\omega^2)^2+\omega^2\gamma^2};\quad \Im G(\omega)=\frac{\omega\gamma}{(\omega_0^2-\omega^2)^2+\omega^2\gamma^2},
\end{eqnarray}
then $\Re G,\Im G$ indeed satisfy Eqs. (\ref{eKKSDOF}). If $\gamma=0$, then the poles of $G(\omega)$ move to the $\Im\omega=0$ axis but $G(\omega)$ is still analytic in $\Im\omega^+$ and Eqs. (\ref{eKKSDOF}) still apply. Taking the limit $\gamma\rightarrow 0^+$ in (\ref{eLorentzRealImag}), we get
\begin{eqnarray}
\label{eLorentzOscillatorGamma0}
G(\omega)=\frac{1}{\omega_0^2-\omega^2}-\frac{i\pi}{2\omega_0}\left[\delta(\omega+\omega_0)-\delta(\omega-\omega_0)\right]
\end{eqnarray}
where $G(\omega)$ is now understood as a distribution. 

\subsection{Causality for Point Functions}\label{subsec:point}
From here on, unless otherwise stated, Fourier transforms will be evaluated for real frequencies $\omega=p$ (see Fig. \ref{fhalfPlane}), and will be represented by symbols like $G(\omega)$ or $G(p)$. Their extensions in the upper half will be evaluated for complex frequency $k=p+is$ ($s>0$), and denoted by symbols like $G(k)$ The latter will be equivalent to the Laplace transform under the cases considered in this paper.

A general problem of concern will be a relation between an input, $f(t)$, and output, $x(t)$, mediated through a response function, $g(t)$. Input-output relationships can be completely arbitrary, $x=g(f)$, but they reduce to a particularly simple form when certain properties are assumed for the relationship \citep{zemanian1965distribution}. Assumptions that $g$ is bijective, linear, and time-invariant are sufficient to ensure that the input-output relationship is in the form of a convolution integral:
\begin{eqnarray}
\label{ConvRelPointFunction}
x(t)=\int_{-\infty}^{\infty}g(t-\tau)f(\tau)d\tau=\int_{-\infty}^{\infty}g(\tau)f(t-\tau)d\tau=g(t)*f(t)
\end{eqnarray}
At this point, we are also going to define inner products which will be useful going forward:
\begin{eqnarray}
\label{inner}
(g(t),f(t))=\int_{-\infty}^{\infty}g(t)f(t)dt
\end{eqnarray}
We are now going to assume that the Fourier transforms exist for $x(t),g(t),f(t)$, and that they are given by $X(\omega),G(\omega),F(\omega)$ respectively. The Fourier transforms are given by $X(\omega)\equiv\mathcal{F}x=(x(t),e^{i\omega t})$ etc. and the inverse Fourier transforms are given by $x(t)\equiv\mathcal{F}^{-1}X=(X(\omega),e^{-i\omega t})$ etc. We will assume that the convolution theorem applies so that Eq. (\ref{ConvRelPointFunction}) implies:
\begin{eqnarray}
\label{ConvThmPointFunction}
X(\omega)=G(\omega)F(\omega)
\end{eqnarray}
An example of a case when the above assumptions would be true is if $x(t),g(t),f(t)$ belong to $L_2$, where $L_p$ is the space of all functions $F(t)$ for which $(\int_{-\infty}^{\infty}|F(t)|^pdt)^{1/p}$ is finite \citep{paley1934fourier}. The $L_2$ space is also called the space of square integrable functions. Assuming that $f(t)$ is square integrable is often related to the physical restriction that the total energy of the system is finite \citep{toll1956causality}. The question of relevance here is: what can we say about $G(\omega)$ in (\ref{ConvThmPointFunction}) based upon certain restrictions on $g(t)$ such as causality? We will see later that causality implies that $G(\omega)$ has an analytic extension in $\Im\omega^+$. This result is related to the intimate relation between Laplace and Fourier transforms \citep{titchmarsh1948introduction}. Due to this result, the implications of causality are intimately connected to the properties of the Hilbert transform, which is itself related to the following problem: given a real function $a(\omega)$, can we find another real function $b(\omega)$ such that $a+ib$ is analytic in $\Im\omega^+$? If this can be done, then $a,b$ are considered Hilbert transform pairs. Since analyticity requires $a+ib$ to satisfy the Cauchy-Riemann equations, the Hilbert transform $b$ of $a$ can be found by solving a boundary value Laplace problem \citep{Oppenheim1998}. Alternatively, with the requirement of analyticity (and some restrictions on $a,b$), the Hilbert transform pairs are given by the following improper integrals \citep{titchmarsh1948introduction,labuda2014mathematics}:
\begin{eqnarray}
\label{HilbertTransform}
a(\omega)=\frac{1}{\pi}\mathcal{P}\int_{-\infty}^{\infty}\frac{b(\omega')}{\omega'-\omega}d\omega';\quad b(\omega)=-\frac{1}{\pi}\mathcal{P}\int_{-\infty}^{\infty}\frac{a(\omega')}{\omega'-\omega}d\omega'
\end{eqnarray}
Here, $\mathcal{P}$ refers to the principal value of the integral. Going back to our original problem, causality of $g(t)$, therefore, implies the analyticity of $G(\omega)$ in $\Im\omega^+$, which implies that $\Re G,\Im G$ constitute a Hilbert transform pair (Eq. \ref{eKKSDOF}). Titchmarsh's theorem formalizes these ideas \citep{titchmarsh1948introduction,hille1935absolute}:
\begin{namedtheorem}[Titchmarsh]\label{tTitchmarsh}
If $G(\omega)$ is square integrable and it fulfills any one of the four conditions below, then it fulfills all other conditions as well:
\begin{itemize}
    \item Inverse Fourier transform of $G(\omega)$ is causal: $g(t)=0,t<0$
    \item If $k=p+is$ then $G(p)$ is the limit, for almost all $p$, as $s\rightarrow 0^+$ of an analytic function $G(k)$ that is holomorphic in $\Im\omega^+$, and square integrable over any line parallel to the real axis.
    \item Plemelj's first formula applies: 
    $$\Re G(\omega)=\frac{1}{\pi}\mathcal{P}\int_{-\infty}^{\infty} \frac{\Im G(\omega')}{\omega'-\omega}d\omega'$$
    
    \item Plemelj's second formula applies: 
    $$\Im G(\omega)=-\frac{1}{\pi}\mathcal{P}\int_{-\infty}^{\infty} \frac{\Re G(\omega')}{\omega'-\omega}d\omega'$$
    
\end{itemize}
\end{namedtheorem}
{As mentioned earlier, the Plemelj formulae are Hilbert transform relations. They can be derived from a set of more general formulae first discovered by Julian Sokhotskii in the 19$^\mathrm{th}$ century \citep{sokhotskii1873definite} by applying them to the real axis. These formulae were rediscovered by Plemelj \citep{Plemelj1908} (subsequently refined by Privalov \citep{privalov1950boundary,privalov1956randeigenschaften}) as the main ingredient in his solution of the Riemann-Hilbert problem, a specialization of which is the problem of analyticity in $\Im\omega^+$. The general formulae are now known as the Sokhotskii-Plemelj formulae, whereas their specialization to the real axis is sometimes referred to simply as the Plemelj formulae.} The Plemelj formulae may be written in a succinct form involving convolutions:
\begin{eqnarray}
\label{ePlemeljConv}
G(\omega)=-\frac{1}{\pi i}\left[G(\omega)*\mathcal{P}\left(\frac{1}{\omega}\right)\right]
\end{eqnarray}
The Plemelj formulae above are also known as Kramers-Kronig relationships and are sometimes expressed in terms of positive frequency values. This is especially helpful when the input and output, $f(t),x(t)$, are physically measurable quantities which are expected to be real. In such a case, the response function, $g(t)$, must also be real. If $g(t)$ is real, then we must have $G(-\omega)=G(\omega)^*$, or that $\Re G$ is an even function of frequency and $\Im G$ is an odd function. For such a case, the integrals in the Plemelj formulae can, instead, be evaluated on the interval $\omega'=[0,\infty)$:
\begin{eqnarray}
\label{eKKPositiveF}
\nonumber \Re G(\omega)=\frac{2}{\pi}\mathcal{P}\int_{0}^{\infty} \omega^{'}\frac{\Im G(\omega')}{\omega^{'2}-\omega^2}d\omega'\\
\Im G(\omega)=-\frac{2}{\pi}\mathcal{P}\int_{0}^{\infty}\omega \frac{\Re G(\omega')}{\omega^{'2}-\omega^2}d\omega'
\end{eqnarray}
 This specialization of the Plemelj formulae to real response functions was first noticed in the context of x-ray scattering by Kronig whose derivation did not involve any mention of complex functions or analyticity, and who only derived one of the two formulae \citep{Kronig1926,Bohren2010}. In that paper, Kronig derived the dispersion relation relating the real and imaginary parts of the complex refractive index for electromagnetic wave propagation. Shortly thereafter, Kramers \citep{kramers1927diffusion} argued that the real and imaginary parts of atomic polarizability are a Hilbert transform pair and derived both dispersion relationships. Gorter and Kronig subsequently argued, again without any reference to causality, that the dispersion relations apply to both magnetic as well as electric cases \citep{Gorter1936}. From the perspective of physics, the earliest derivations of the dispersion relations, therefore, did not have their origins in complex analysis, analyticity, or causality, but rather in the analysis of the Sellmeier equation (generalized form of Eq. \ref{eSDOF}). In this period, causality was treated in the context of prohibition of faster than light travel by Sommerfeld \citep{Sommerfeld1914} and Brillouin \citep{Brillouin1914}, but without any reference to dispersion relations. It was Kronig \citep{kronig1942} who first connected the dispersion relations to causality by invoking arguments based on the necessary analytic continuation of the Fourier transforms of causal transfer functions in $\Im\omega^+$. Titchmarsh finally presented a set of proofs which crystallized the results in the form of the Titchmarsh's theorem presented above \citep{titchmarsh1948introduction}. At this point, it was clear that the Titchmarsh theorem applies to any causal transfer function whose Fourier transform was square integrable, thus connecting the real and imaginary parts of such a Fourier transform.

The Titchmarsh's theorem depends upon the assumption that $G(\omega)$ is square integrable. However, a response function can be causal without its Fourier transform satisfying the square integrability assumption. What dispersion relations can be derived in such a case? Toll \citep{toll1956causality} considered a specific manifestation of such a problem where $x(t)$ and $f(t)$ were taken as square integrable, but in such a way that $\int_{-\infty}^{\infty}|x(t)|^2dt\leq A\int_{-\infty}^{\infty}|f(t)|^2dt$, where $A$ is a constant. Such a condition may represent a system in which the output energy is at most equal to the input energy. For this system, it can be shown that $\vert G(\omega)\vert^2\leq A$ and, therefore, $G(\omega)$ is not square integrable but merely bounded. For the case when $G(\omega)$ is merely bounded by a constant, we can define another function \citep{hilgevoord1960,nussenzveig1972causality}:
\begin{eqnarray}
\label{eSubtraction1}
D(\omega,\omega_0)=\frac{G(\omega)-G(\omega_0)}{\omega-\omega_0}
\end{eqnarray}
where $\omega_0$ is an arbitrarily chosen value on the real axis. If we assume that $D(\omega,\omega_0)$ is differentiable at $\omega_0$, then it is bounded as $\omega\rightarrow \omega_0$. Furthermore, $D(\omega,\omega_0)$ is analytic in $\Im\omega^+$ due to the analycity of $G(\omega)$, and it is square-integrable on the real axis due to the bounded nature of $G(\omega)$. It can be shown that $D(\omega,\omega_0)$ is, in fact, a causal transform and it, therefore, satisfies Titchmarsh's theorem. We can write Plemelj formulae for $D(\omega,\omega_0)$, which can be simplified to produce modified dispersion relations for $G(\omega)$:
\begin{eqnarray}
\label{ePlemeljSubtraction1}
\nonumber \Re G(\omega)=\Re G(\omega_0)+\frac{\omega-\omega_0}{\pi}\mathcal{P}\int_{-\infty}^{\infty} \frac{\Im[G(\omega')-G(\omega_0)] }{\omega'-\omega_0}\frac{d\omega'}{\omega'-\omega}\\
\Im G(\omega)=\Im G(\omega_0)-\frac{\omega-\omega_0}{\pi}\mathcal{P}\int_{-\infty}^{\infty} \frac{\Re[G(\omega')-G(\omega_0)] }{\omega'-\omega_0}\frac{d\omega'}{\omega'-\omega}
\end{eqnarray}
More generally, dispersion relations with additional subtractions can always be written in the following way \citep{nussenzveig1972causality}:
\begin{eqnarray}
\label{ePlemeljSubtractionN}
G(\omega)=\bar{G}(\omega,\omega_0)+\frac{(\omega-\omega_0)^{n+1}}{\pi i}\mathcal{P}\int_{-\infty}^{\infty} \frac{[G(\omega')-\bar{G}(\omega',\omega_0)] }{(\omega'-\omega_0)^{n+1}}\frac{d\omega'}{\omega'-\omega}
\end{eqnarray}
where
\begin{eqnarray}
\bar{G}(\omega,\omega_0)=G(\omega_0)+\sum_{i=1}^{n}\frac{(\omega-\omega_0)^n}{n!}G^{(n)}(\omega_0)
\end{eqnarray}
In the above, we have assumed that derivatives of $G(\omega)$ up to order $n+1$ exist at $\omega=\omega_0$. The above relations are valid if $G(\omega)=O(\omega^k),k\leq n$. In summary, if $g(t)$ is causal but $G(\omega)$ is not square-integrable but exhibits polynomial growth, modified dispersion relations may still be derived for $G(\omega)$ by considering appropriate subtractions. Successively higher orders of subtractions allow us to treat $G(\omega)$ with successively weaker integrability requirements. 

\subsection{Causality for distributions}\label{subsec:distributions}
The above analysis can be considerably unified by considering it under the theory of distributions. Furthermore, assuming that the inputs, outputs, and the Green's functions are ordinary functions is too restrictive for many physical applications. Here, we present only the salient ideas and refer the reader to more extensive texts for details on distribution theory and operator theory \citep{zemanian1965distribution,beltrami1966,livshits1973operators,gohberg1978introduction}. Distributions are defined as linear functionals which operate on a set of test functions through the inner production operation. For example, the $\delta(t)$-function is a distribution which is defined by its action on test functions $\phi(t)$:
\begin{eqnarray}
(\delta,\phi)=\int_{-\infty}^{\infty}\delta(t)\phi(t)dt=\phi(0)
\end{eqnarray}
The integral has to be well-defined, therefore, the class to which the test functions belong places restrictions on the class to which the distributions can belong. Therefore, consistent definitions of some test function spaces and corresponding distribution spaces are needed. We define the set $\mathcal{D}$ of all test functions which are infinitely differentiable (belong to the class $C^\infty$), and have compact support. We define the space of Schwartz distributions $\mathcal{D}'$ as containing distributions which act on the space of test functions belonging to $\mathcal{D}$ through the inner product $(.,.)$, and produce complex numbers. Schwartz distributions with support only in $[0,\infty)$ belong to $\mathcal{D}_+'$, with $\mathcal{D}_+'\subset\mathcal{D}'$. We would like to define the Fourier transform of a distribution using the operation:
\begin{eqnarray}
(\mathcal{F}T,\phi)=(T,\mathcal{F}\phi)
\end{eqnarray}
where $\mathcal{F}\phi$ is the well known Fourier transform of a function $\phi$ (Eq. \ref{eFourierPoint}). However, if $\phi\in\mathcal{D}$, then the above is not allowed, since $\mathcal{F}\phi$ can be shown to not belong to $\mathcal{D}$. To define Fourier transforms of distributions, therefore, we need a different space of test functions. We define $\mathcal{L}$ as the space of rapidly decreasing test functions characterized by $\phi(t)\in C^\infty$ which, together with all their derivatives, decrease faster than any inverse power of $t$ as $\vert t\vert\rightarrow \infty$:
\begin{eqnarray}
\lim_{\vert t\vert\rightarrow\infty}\vert t^p\frac{\partial^m \phi(t)}{\partial t^m}\vert=0;\quad p,m=0,1,...
\end{eqnarray}
$\mathcal{D}\subset\mathcal{L}$ due to compact support in $\mathcal{D}$. An example of a function which is in $\mathcal{L}$ but not in $\mathcal{D}$ is $\exp(-t^2)$. We can show that if $\phi\in\mathcal{L}$, then $\mathcal{F}\phi\in\mathcal{L}$, and we can use this property to define a set of distributions whose Fourier transforms can be evaluated. We define the class of temperate distributions, $\mathcal{L}'$, as the set of distributions which are linear functionals on $\mathcal{L}$. Now we can define the Fourier transform of a distribution $T\in\mathcal{L}'$ through the relation $(\mathcal{F}T,\phi)=(T,\mathcal{F}\phi)$, and note that $\phi,\mathcal{F}\phi\in\mathcal{L}$, $T,\mathcal{F}T\in\mathcal{L}'$, and $\mathcal{L}'\subset\mathcal{D}'$. We will consider the implications of causality for transfer functions which are in $\mathcal{L}'$. As some examples, $\mathcal{L}'$ contains all functions of polynomial growth and all distributions of bounded support. The former means that all functions in $L_p$ belong to the space of temperate distributions, and the latter means that the $\delta$ function also belongs to this space (along with all its derivatives). Also relevant here, all the point functions considered above, for which the method of subtractions was employed, belong to $\mathcal{L}'$.

Now consider the input-output problem in the following general form:
\begin{eqnarray}
x(t)=g(f(t))
\end{eqnarray}
where both the input, $f(t)$, and the output, $x(t)$, are considered to be distributions in $\mathcal{D}'$. The connection between input and output is considered to be an arbitrary operator $g$. If $g$ is linear, time-translation invariant, and continuous, then the above relation can be shown to be a convolution operation \citep{zemanian1965distribution}:
\begin{eqnarray}
x(t)=g(t)*f(t)
\end{eqnarray}
where $g(t)\in\mathcal{D}'$, and convolution is considered in the sense of distributions \citep{nussenzveig1972causality}. If we now further restrict the quantities involved so that $f(t),x(t),g(t)\in\mathcal{L}'$, then it ensures that the following Fourier transforms exist:
\begin{eqnarray}
F(\omega)=\mathcal{F}f(t),\quad X(\omega)=\mathcal{F}x(t),\quad G(\omega)=\mathcal{F}g(t),
\end{eqnarray}
that they belong to $\mathcal{L}'$, and that the following relations are satisfied by them:
\begin{eqnarray}
X(\omega)=G(\omega) F(\omega).
\end{eqnarray}
Streater-Wightman \citep{Streater1964}, Beltrami-Wohlers \citep{beltrami1965distributional,beltrami1966distributionalb}, and Lauwerier \citep{lauwerier1962hilbert}, independently showed that if $g(t)\in\mathcal{L}_+^{'}$ (causality condition), then its Fourier transform $G(\omega)$ (which is in $\mathcal{L}^{'}$) has an analytical continuation in the upper half of the complex plane. This analytical continuation, $G(k)$, is the Laplace transform of $g_t$. With $k=p+is$, the Laplace transform is defined through the Fourier transform itself using $G(k)=\mathcal{F}(g(t)e^{-st})$, with $p$ playing the role of the real frequency $\omega$. This is the distributional analogue of the point function result which connects causality to analyticity in the upper half of the complex plane. Beltrami and Wohlers \citep{beltrami2014distributions} showed that if one uses this analytical continuation, then dispersion relations can be derived connecting the real and imaginary parts of $G(\omega)$ without any subtraction terms even for those distributions which show polynomial growth (i.e. tempered distributions). Note the contrast with Eq. (\ref{ePlemeljSubtractionN}) where subtraction terms appear in the dispersion relations of $G(\omega)$ when it shows polynomial growth. If one does not use analytic continuation, then dispersion relations can still be derived but they will contain subtraction terms \citep{nussenzveig1972causality}. G\"uttinger \citep{guttinger1966generalized} showed that this discrepancy is connected to the fact that the product of two distributions is generally determined only up to an arbitrary distribution. When analytic continuation is considered, then the generalized Hilbert transform can be written for a distribution $G(\omega)$ in $\mathcal{L}'$ \citep{beltrami1966distributionalb,waters2000application}:
\begin{eqnarray}
\label{eHilbertGen}
G(\omega)=-\frac{\omega^n}{\pi i}\left[\frac{G(\omega)}{\omega^n}*\mathcal{P}\left(\frac{1}{\omega}\right)\right]
\end{eqnarray}
where, as before, $*$ represents convolution, and $\mathcal{P}$ is the principal value. In the above, $n$ is any integer for which $\mathcal{F}^{-1}G(\omega)=g(t)=D^nu_0$. $u_0$ is some tempered distribution which is locally square integrable everywhere, and $D^n$ represents the $n^\mathrm{th}$ derivative. The special case of $n=0$ corresponds to a situation where $G(\omega)$ belongs to a space $\mathcal{D}_{L_2}'$ \citep{beltrami2014distributions} which encompasses the set of all square integrable functions ($L_2$ functions). This means that as a special case, if $G(\omega)\in L_2$, then the generalized Hilbert transform reduces to:
\begin{eqnarray}
G(\omega)=-\frac{1}{\pi i}\left[G(\omega)*\mathcal{P}\left(\frac{1}{\omega}\right)\right]
\end{eqnarray}
If $G(\omega)$ is an ordinary function, then the above is the same expression as Eq. (\ref{ePlemeljConv}). The generalized Hilbert transform can be used to relate the real and imaginary parts of $G(\omega)$:
\begin{eqnarray}
\label{eKramersKronigDist}
\Re G(\omega)=-\frac{\omega^n}{\pi}\left[\frac{\Im G(\omega)}{\omega^n}*\mathcal{P}\left(\frac{1}{\omega}\right)\right];\quad 
\Im G(\omega)=\frac{\omega^n}{\pi}\left[\frac{\Re G(\omega)}{\omega^n}*\mathcal{P}\left(\frac{1}{\omega}\right)\right]
\end{eqnarray}

The integer $n$ is indicative of the differentiability properties of $g(t)$, and also the growth properties of its Fourier transform $G(\omega)$. To make these more explicit, we summarize certain important distributional results. The main idea is a representation theorem \citep{beltrami2014distributions} according to which every distribution in $\mathcal{L}'$ is the $n^\mathrm{th}$ order derivative of a distribution $u_0$. $g(t)=D^nu_0$ ensures that its Fourier transform can be written as $G(\omega)=(i\omega)^nv(\omega)$, where $v(\omega)=\mathcal{F}u_0$. As an illustrative example, one may consider (for $n=0$), $G(\omega)$ to be an $L_2$ function, which would immediately take us to the Titchmarsh's theorem for point functions mentioned earlier. For $n>0$, however, $G(\omega)=(i\omega)^nv(\omega)$ represents polynomial growth of the Fourier transform in the spirit of the cases of subtractions considered for point functions. For such cases, the generalized Hilbert transform (Eq. \ref{eHilbertGen}) and the associated generalized dispersion relations are immediately applicable. The integer $n$, therefore, represents the order of differentiability which connects $g(t)$ to the space of locally integrable tempered functions, the order of polynomial growth in $\omega$ which exists in $G(\omega)$, and eventually determines the precise generalized dispersion relations which connect $\Re G(\omega)$ and $\Im G(\omega)$.

\subsection{Passivity and Causality}\label{subsec:passcaus}
For a physical system with an input output relation in the convolution form: $x(t)=g(t)*f(t)$, the requirement that the system be passive (output energy cannot exceed input energy) automatically implies that the system is causal as well. To understand this inter-relationship, we first reiterate that the causality requirement is $g(t)\in\mathcal{D}_+$ and note that it means that if $f(t)=0$ for $t<t_0$, then it implies that $x(t)=0$ for $t<t_0$ as well. It will be seen later that for scattering problems, passivity can be framed in the following two forms (scattering formalism):
\begin{subequations}
\label{ePassive}
\begin{eqnarray}
\int_{-\infty}^{\infty}\left(|f(t)|^2-|x(t)|^2\right)dt\geq0\\
\int_{-\infty}^{t}\left(|f({t'})|^2-|x({t'})|^2\right)d{t'}\geq0,\;\forall\;t
\end{eqnarray}
\end{subequations}
Eq. (\ref{ePassive})b implies causality whereas Eq. (\ref{ePassive})a requires the additional assumption of causality \citep{guttinger1966generalized}. The passivity relations can be framed in another set of forms, called the immittance form, which emerges naturally in certain problems. The introduction of new variables $v(t)=f(t)+x(t)$ and $j(t)=f(t)-x(t)$ allows us to write the passivity conditions as:
\begin{subequations}
\label{ePassiveiv}
\begin{eqnarray}
\Re\int_{-\infty}^{\infty}v(t)^*j(t)dt\geq0\\
\Re\int_{-\infty}^{t}v({t'})^*j({t'})d{t'}\geq0,\;\forall\;t
\end{eqnarray}
\end{subequations}
To show that Eq. (\ref{ePassiveiv})b implies causality, we consider arbitrary real $j,j_0$ with $j=0,t<t_0$, and an input to the system $j_1=j_0+\alpha j$ which produces a corresponding real output $v_1=v_0+\alpha v$. Eq. (\ref{ePassiveiv})b implies that $\int_{-\infty}^{t}v_1j_1d{t'}\geq 0$ for all $t$. Since for $t<t_0$ we have $j=0$, this integral for $t<t_0$ implies $\int_{-\infty}^{t}v_0j_0d{t'}+\alpha\int_{-\infty}^{t}vj_0d{t'}\geq 0$. Since $\alpha$ is arbitrary, this can only be true if $\int_{-\infty}^{t}vj_0d{t'}=0$ for all $t<t_0$. Furthermore, since $j_0$ is arbitrary, this can only be true if $v(t)=0$ for all $t<t_0$. Therefore, $j(t),v(t)$ are simultaneously zero for $t<t_0$ which means that $x(t),f(t)$ must also be simultaneously zero for $t<t_0$. Therefore, the passivity requirement (Eq. \ref{ePassive}b or \ref{ePassiveiv}b) automatically implies causality. While these results were provided in the condensed form above by G\"uttinger \citep{guttinger1966generalized}, there were earlier contributions in the area of passive network theory which laid the foundation of some of these ideas, especially those dealing with the passivity of the system. These early papers in network theory also dealt with tensorial transfer functions which characterized the passive networks.

\section{Passivity from Scattering and Immittance perspectives}\label{sec:scattImm}

{Before discussing the connection between causality and passivity, it is useful to clarify a point of convention. While in various areas of physics, it is customary to talk about the analyticity of some function $Z(k)$ in the upper half in connection with causality, this convention is not customary in network theory and control theory. In these latter fields, it is common to talk about the analyticity of $Z(k)$ in the right half, which indicates stability of the system (especially in control theory) -- a concept closely related to causality. The difference between the two conventions rests on how Laplace transform is being defined. For a point function $\phi(t)$, real $\omega,s$, and $k=\omega+is,\hat{k}=s+i\omega$, two relevant definitions of Laplace transform are $Z(k)=(\phi(t),e^{ikt})$ and $\hat{Z}(\hat{k})=(\phi(t),e^{-\hat{k}t})$. The former has a region of convergence (if it exists) in the upper half of $k$, whereas the latter has it in the right half of $\hat{k}$. To keep the discussions on passivity consistent with history and context, it will be implicitly understood that the right half picture is being referred to in this section and the next. However, it should be noted that both the upper and right halves are denoted by $s>0$.} 

\begin{figure}[htp]
\centering
\includegraphics[scale=.9]{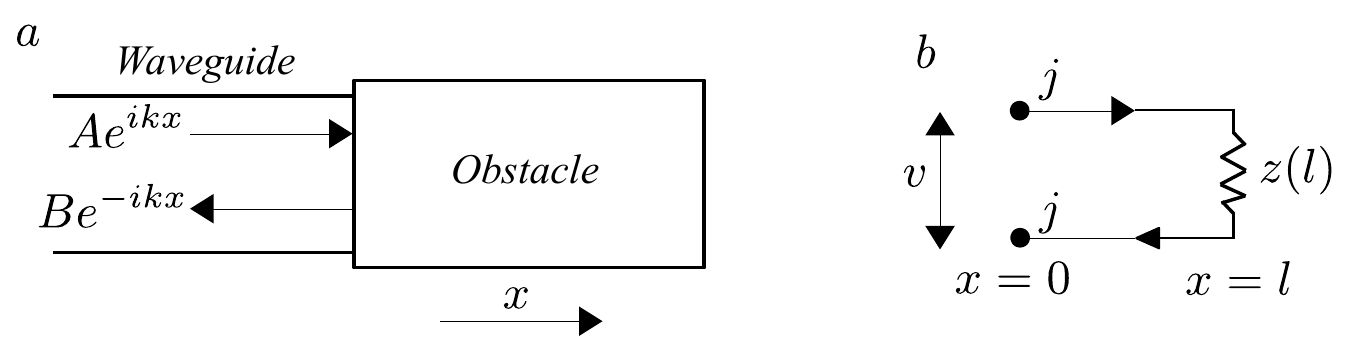}
\caption{Scattering and Immittance perspectives for a simple network.}\label{fScattImm}
\end{figure}

In general, the input-output relations may be written either in a form which involves $x(t),f(t)$, or in a form which involves $v(t),j(t)$. The primary distinction between the two cases is the statement of the passivity condition. In the former case, the passivity conditions are of the form given in Eqs. (\ref{ePassive}) and the system is said to be in a scattering form. In the latter case, the passivity conditions are Eqs. (\ref{ePassiveiv}) and the system is said to be in an immittance form. A physical example which illustrates this distinction comes from the early work in network theory \citep{belevitch1962summary}. Fig. (\ref{fScattImm}) shows the schematics of a very simple electrical network in the scattering and immittance forms. Fig. (\ref{fScattImm})a shows a waveguide which is terminated at an obstacle. An incident harmonic wave, $A(t)e^{ikx}$, traveling in the $+x$ direction is converted into a reflected harmonic wave $B(t)e^{-ikx}$ after interacting with the obstacle. The obstacle is passive which means that the energy in the reflected wave may not be larger than the energy in the incident wave. With appropriate normalization, the energy in the incident (reflected) wave can be shown to be equal to $|A(t)|^2$ ($|B(t)|^2$). Therefore, the passivity statements in the scattering form become:
\begin{subequations}
\label{ePassiveScatt}
\begin{eqnarray}
\int_{-\infty}^{\infty}\left(|A(t)|^2-|B(t)|^2\right)dt\geq0\\
\int_{-\infty}^{t}\left(|A({t'})|^2-|B{(t')})|^2\right)d{t'}\geq0,\quad\forall\;t
\end{eqnarray}
\end{subequations}
For electrical networks, these waves may represent waves of electrical and magnetic fields in waveguides, which implies that they also represent waves of voltage and current. To be more specific, if the waveguide is a coaxial cable, then under appropriate low frequency limits, the energy is transmitted in the \emph{TEM}-mode with nonzero electric field component $E_r$, and nonzero magnetic field component $H_\phi$ \citep{montgomery1948912}. $E_r,H_\phi$ satisfy Maxwell's equations whose solutions are left and right traveling waves of the form discussed here. Since the current is linearly related to the magnetic field and the voltage is linearly related to the electric field, the $TEM$-mode wave solutions in the coaxial waveguide can be transformed into waves of voltage and current. {Waveguides composed of two or more unconnected conducting elements (such as a coaxial cable)}, can support \emph{TEM} modes and, therefore, can behave as transmission lines in the low frequency limit. One can either study such problems from a purely Maxwellian perspective or from a simpler transmission line perspective, which is eventually based upon the Maxwellian perspective under appropriate frequency limits. The Maxwellian perspective, in the present instance, involves solving the Maxwell's equations of motion for the 3-D electromagnetic field subject to the boundary conditions imposed on the surfaces and terminals of the coaxial cable. The transmission line perspective, on the other hand, involves solving a set of 1-D partial differential equations in terms of voltage and current subject to appropriate impedance boundary condition, which represent the effect of the obstacle (represented as $z(l)$ in Fig. \ref{fScattImm}b):
\begin{subequations}
\label{eTransmissionLine}
\begin{eqnarray}
\frac{\partial v}{\partial x}=-\bar{z}j;\quad \frac{\partial j}{\partial x}=-\bar{y}v
\end{eqnarray}
\end{subequations}
Here, $\bar{z},\bar{y}$ are the series impedance and shunt admittance per unit length of the line. The general solutions to the above are left and right traveling waves, but these are waves of voltage or current. If voltage is chosen as the primary wave, then the voltage at any location $x$ will be proportional to $A(t)e^{ikx}+B(t)e^{-ikx}$, whereas the current will be proportional to $A(t)e^{ikx}-B(t)e^{-ikx}$. To simplify further and without any loss of generality, at a specific location $x=0$, the voltage in the transmission line is proportional to $A(t)+B(t)$ whereas the current is proportional to $A(t)-B(t)$. These are precisely the kinds of transformations which were discussed in the last section as we transformed $x,f$ to $v,j$. Since energy at a point in an electrical circuit is equal to $\Re v^*j$, the passivity statements in the immittance form become:
\begin{subequations}
\label{ePassiveImm}
\begin{eqnarray}
\Re\int_{-\infty}^{\infty}v(t)^*j(t)dt\geq0\\
\Re\int_{-\infty}^{t}v({t'})^*j({t'})d{t'}\geq0\quad\forall\;t
\end{eqnarray}
\end{subequations}
In analogy with the last section, it is possible to relate the scattering amplitudes thorough a convolution operation $B(t)=g(t)*A(t)$, where $g(t)$ is called the scattering coefficient. $v(t),j(t)$ can be evaluated at various points in the transmission line. Particularly, they can be evaluated at the terminal ends ($x=0$ in Fig. \ref{fScattImm}b), and related to each other through impedance and admittance convolutions. Specifically, $v(t)=z(t)*j(t)$ and $j(t)=y(t)*v(t)$. Therefore, the link between the wave nature of the problem admitting a scattering description and a transmission line nature of the problem admitting an immittance description is complete for the very simple case shown in Fig. (\ref{fScattImm}). In network analysis, while the immittance descriptions were found appropriate for low frequency applications, the scattering descriptions were found more naturally suitable for higher frequency applications such as networks operating in the microwave regime \citep{montgomery1948912}. In either case, network theory provides the direct means for associating with an electrical network a mathematical description which characterizes the behavior of that network \citep{mcmillan1952introduction,bode1945network}. 

\begin{figure}[htp]
\centering
\includegraphics[scale=.25]{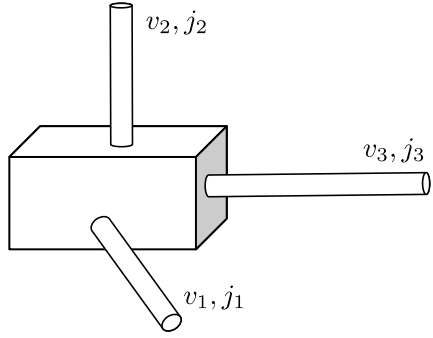}
\caption{Schematic of a three-terminal pair junction or a 3-port network}\label{fThreeTerminal}
\end{figure}

The problem described in Fig. (\ref{fScattImm}) is scalar in nature as the transfer functions $g,z,y$ are scalar quantities which must individually satisfy requirements of passivity and causality \citep{foster1924reactance,brune1931synthesis,bott1949impedance}. Such a network, with only one transmission line, has been variously referred to as a single terminal pair network or a 1-port network. For such a scalar network, Raisbeck \citep{raisbeck1954definition} showed that passivity implies that $Z(k),Y(k)$ are analytic in the region $s>0$, and that they are real for $p=0$ because the transfer functions were assumed real. Furthermore, $\Re Z(k),\Re Y(k)\geq 0$ for $s\geq 0$. Raisbeck only considered network representations in the immittance form and did not consider the quantities as distributions but we are continuing to use the same notations introduced earlier for consistency. The ideas from the single transmission line network can now be extended to a network consisting of multiple transmission lines connected through a junction (schematic of a three terminal pair network or a 3-port network is shown in Fig. \ref{fThreeTerminal}). For the case of $n$ transmission lines, there are $n$ voltages and $n$ currents which could be measured at the available terminal pairs. Therefore, voltage and currents can be represented as vectors $\mathbf{v,j}$, each with $n$ elements. Similarly, there are $n$ incident waves and $n$ emergent waves, and their amplitudes can be similarly represented by n-dimensional vectors $\mathbf{A,B}$. From here on, the scattering amplitude vectors $\mathbf{A,B}$ will instead be represented by $\mathbf{f,x}$, respectively, to maintain continuity and consistency with previous discussions. The transfer functions now become square matrices of size $n\times n$ given by symbols $\mathbf{z,y,g}$, which are the impedance, admittance, and scattering matrices respectively \citep{bayard1949synthese,mcmillan1952introduction,oono1950synthesis}. Such higher dimensional networks are sometimes called $n-$port networks.

Raisbeck generalized the 1-port arguments to a general n-port case (but only for networks with an immittance representation), where a vector of voltages $\mathbf{v}(t)\in \mathbb{R}^{n\times 1}$ is related to a vector of currents $\mathbf{j}(t)\in \mathbb{R}^{n\times 1}$ through matrices of real impedance and admittance transfer functions $\mathbf{z}(t),\mathbf{y}(t)\in\mathbb{R}^{n\times n}$. For simplicity, this relation is expressed as $\mathbf{v}=\mathbf{z}*\mathbf{j},\mathbf{j}=\mathbf{y}*\mathbf{v}$ where $*$ indicates a convolution operation in time as well as a matrix multiplication operation on the indices:
\begin{eqnarray}
\label{eMatrixConvolution} 
\mathbf{v}=\mathbf{z}*\mathbf{j}: v_l(t)=z_{lm}(t)*j_m(t);\quad l,m=1,2...n
\end{eqnarray}
Elementwise Fourier (Laplace) transforms of $\mathbf{z}(t),\mathbf{y}(t)$ are given by impedance $\mathbf{Z}(\omega)$ ($\mathbf{Z}(k)$) and admittance $\mathbf{Y}(\omega)$ ($\mathbf{Y}(k)$) matrices, with the system relations $\mathbf{V}=\mathbf{Z}\mathbf{J},\mathbf{J}=\mathbf{Y}\mathbf{V}$, where the capital letters denote the transformed quantities (Fourier or Laplace), and regular matrix multiplication is implied. 

Raisbeck showed that passivity condition, $\int_{-\infty}^{\infty}\mathbf{v}(t)^\dagger\cdot\mathbf{j}(t)dt\geq 0$, implies that the hermitian parts of $\mathbf{Z},\mathbf{Y}$, given by
\begin{eqnarray}
\label{eHermitian} 
\mathbf{Z}^h=\frac{1}{2}\left[\mathbf{Z}+\mathbf{Z}^\dagger\right];\quad \mathbf{Y}^h=\frac{1}{2}\left[\mathbf{Y}+\mathbf{Y}^\dagger\right],
\end{eqnarray}
are positive definite. In the above, $\dagger$ represents a conjugate transpose operation. Raisbeck's analysis assumed an immittance form of the passivity definition similar to Eq. (\ref{ePassiveiv})a, which necessitated the additional assumption of causality. It is important to note that causality does not automatically follow from the passivity definition that Raisbeck assumed and, therefore, the positive definiteness properties of $\mathbf{Z}^h,\mathbf{Y}^h$ do not automatically imply any dispersion relations which one might derive from causality. Connection between passivity and causality, however, was not the primary concern of Raisbeck in any case.

To build upon Raisbeck's work, Youla et al. \citep{youla1959bounded} assumed a definition of passivity similar to Eq. (\ref{ePassiveiv})b, and considered the tensorial problem from the perspective of the scattering matrix. They considered $\mathbf{x}(t),\mathbf{f}(t),\mathbf{v}(t),\mathbf{j}(t)$ to be in $L_{2_n}$, which is the space of vectors with elements which are in $L_2$, and they took the passivity condition in its immittance form as:
\begin{eqnarray}
\label{ePassiveYoula} 
\Re\int_{-\infty}^{t}\mathbf{v}(t')^\dagger \mathbf{j}(t')dt'\geq 0
\end{eqnarray}
This definition also trivially implies that the following relation is also true:
\begin{eqnarray}
\label{ePassiveYoulaRaisbeck} 
\Re\int_{-\infty}^{\infty}\mathbf{v}(t)^\dagger \mathbf{j}(t)dt\geq 0
\end{eqnarray}
which is the passivity definition assumed by Raisbeck. This relation also ensures that the results in Eq. (\ref{eHermitian}) are still valid. We have also seen earlier that the passivity definition (\ref{ePassiveYoula}) automatically implies causality, which poses some further conditions on $\mathbf{Z}(k),\mathbf{Y}(k),\mathbf{Z}(\omega),\mathbf{Y}(\omega)$. Youla et al. showed that $\mathbf{Z}(\omega),\mathbf{Y}(\omega)$ are bounded $n\times n$ matrices whose individual elements are the $L_{2}-$Fourier transforms of the corresponding elements of $\mathbf{z,y}$. The individual elements of the Laplace transform matrices, $\mathbf{Z}(k),\mathbf{Y}(k)$, are analytic and uniformly bounded in $s>0$, and take the Fourier transform matrices, $\mathbf{Z}(\omega),\mathbf{Y}(\omega)$, as their boundary values in the limit $s\rightarrow 0$. 

Furthermore, as mentioned earlier, $\mathbf{f,x}$ (incident and emergent scattering coefficients respectively), are linearly related to each other through the scattering matrix transfer function: $\mathbf{x}(t)=\mathbf{g}(t)*\mathbf{f}(t)$ which, in the transform domain, becomes $\mathbf{X}(k)=\mathbf{G}(k)\mathbf{F}(k)$. $\mathbf{G}(k)$ is known as the scattering matrix \citep{gross1941theory,kronig1942,kronig1946supplementary}. Youla et al. showed that as a consequence of passivity and causality, the individual elements of the scattering matrix $\mathbf{G}(k)$ are analytic in $s>0$, and uniformly bounded for $s\geq 0$. Furthermore, the passivity conditions also mean that if we define a $n\times n$ matrix:
\begin{eqnarray}
\label{eQ} 
\mathbf{Q}(k)=\mathbf{1}_n-\mathbf{G}^\dagger(k)\mathbf{G}(k),
\end{eqnarray}
then $\mathbf{Q}(k)$ can be shown to be non-negative definite for all $s\geq 0$. Specifically, for general $\mathbf{b}$:
\begin{eqnarray}
\label{eQPositiveDefinite} 
\mathbf{b}^\dagger\mathbf{Q}(k)\mathbf{b}\geq 0;\quad s\geq 0
\end{eqnarray}

Youla et al. \citep{youla1959bounded}, in their paper, assumed that the field variables $\mathbf{x}(t),\mathbf{f}(t),\mathbf{i}(t),\mathbf{j}(t)$ were measurable functions. Zemanian improved upon this by bringing the analysis of single-valued $n-$ports \citep{konig1958lineare} under the umbrella of distribution theory for the first time \citep{zemanian1963n}. While Youla et al. considered the problem essentially from a scattering perspective, Zemanian considered it from the immittance perspective. Wohlers and Beltrami \citep{wohlers1965distribution}, and Beltrami \citep{beltrami1967linear} finally discussed the two approaches within a unified distributional framework. Here, we discuss the salient results from this unified perspective. These results bring the scattering and immittance frameworks together, and unify passivity results with dispersion results within a tensorial and distributional framework.

\section{Passivity and Causality from a Distributional and Tensorial Perspective}\label{sec:Passive}
A tensor of distributions $\mathbf{f}(t)$ is defined through its actions on a test function $\phi(t)$, both in appropriate spaces. Specifically, $\langle\mathbf{f}(t),\phi(t)\rangle$ is the matrix of complex numbers obtained by replacing each element of $\mathbf{f}(t)$ by the number that this element assigns to the testing function $\phi(t)$ through the inner product operation. Zemanian introduced tensorial distribution spaces to admit tensors of distributions of appropriate ranks. For example, $\mathcal{D}^{'}_{n\times n\times n\times n}$ is the space of all fourth order tensors whose elements are distributions in $\mathcal{D}^{'}$ etc. Zemanian showed that a single-valued, linear, time-invariant, and continuous input output relation can be written in the convolution form, $\mathbf{v}=\mathbf{z}*\mathbf{j}$, where $\mathbf{v,z,j}$ are tensors of distributions in appropriate spaces, and $*$ denotes a convolution in time as well as appropriate tensorial contraction (see Eq. \ref{eMatrixConvolution}). Causality is understood in the usual sense either through the statement that $\mathbf{j}(t)=0;t<t_0$ implies $\mathbf{v}(t)=0;t<t_0$, or through the requirement that $\mathbf{z}(t)\in\mathcal{D}^{'}_{n\times n+}$ (if $\mathbf{z}(t)$ is a matrix of distributions). The Fourier and Laplace transforms of $\mathbf{z}(t)$, given by $\mathcal{F}\mathbf{z},\mathcal{L}\mathbf{z}$, are defined by taking the distributional Fourier and Laplace transforms of the individual elements of $\mathbf{z}$. The Fourier and Laplace transforms will also be represented by $\mathbf{Z}(\omega),\mathbf{Z}(k)$, respectively, in accordance with earlier established conventions.

Beltrami \citep{beltrami1967linear} identified both scattering and immittance problems for tensorial and distributional input-output relationships, and the rest of the discussion in this section follows closely from that paper. We assume that $\mathbf{x,f,j,v}$ are vectors of distributions with appropriate dimensions and in appropriate distributional spaces. A convolutional scattering relationship exists between $\mathbf{x,f}$ through a matrix of distributions in appropriate spaces: $\mathbf{x}=\mathbf{g}*\mathbf{f}$, and an immttance relationship exists between $\mathbf{v,j}$ through matrices of distributions in appropriate spaces: $\mathbf{v}=\mathbf{z}*\mathbf{j};\mathbf{j}=\mathbf{y}*\mathbf{v}$. The following theorems summarize the connections between causality, passivity, and the resulting dispersion relations for matrix valued distributional transfer functions in scattering or immittance forms \citep{beltrami1967linear}.

\begin{theorem}\label{tCausality}
If $\mathbf{w}$ is a matrix of distributional transfer functions corresponding to a linear and causal system, then
\begin{itemize}
    \item Each element of $\mathbf{W}(k)$ is analytic for $s>0$, and the Laplace transform, $\mathbf{W}(k)$, has the Fourier transform, $\mathbf{W}(\omega)$, as its boundary value as $s\rightarrow 0$.
    \item For some integer $m\geq0$, and for all $n\geq m$ 
\begin{eqnarray}
\label{eHilbertGenDis}
\mathbf{W}(\omega)=-\frac{\omega^n}{\pi i}\left[\frac{\mathbf{W}(\omega)}{\omega^n}*\mathcal{P}\left(\frac{1}{\omega}\right)\right]
\end{eqnarray}
\end{itemize}
\end{theorem}

The above theorem encompasses the analogue of the distributional Hilbert transform discussed in Eq. (\ref{eHilbertGen}). It should be noted that the above does not have any subtraction constants which are present in Eqs. (\ref{ePlemeljSubtraction1},\ref{ePlemeljSubtractionN}). This matter is discussed more in detail later but it suffices to say here that the absence of the subtraction constants has to do with the fact that Beltrami has assumed analytic continuation \citep{guttinger1966generalized}. If one does not use analytic continuation, then the dispersion relations above will have subtraction constants \citep{nussenzveig1972causality}. The dispersion relations for the indvidual elements of $\mathbf{W}(\omega)$ follow from these in a manner completely analogous to the scalar distributional case discussed earlier. At this point we can list the requirements posed by causality and passivity for systems in scattering and immittance forms. For transfer functions in scattering form, we have:

\begin{theorem}\label{tCPScatt}
If $\mathbf{g}$ is a matrix of distributional transfer functions corresponding to a linear, passive, and causal system in the scattering form, with its elementwise distributional transforms given by $\mathbf{G}(\omega),\mathbf{G}(k)$, then all of the following are true for $\mathbf{G}(\omega)$
\begin{itemize}
    \item $\mathbf{G}^*(\omega)=\mathbf{G}(-\omega)$
    \item $\mathbf{Q}(\omega)=\mathbf{1}_n-\mathbf{G}^\dagger(\omega)\mathbf{G}(\omega)$ is non-negative definite
    \item If the system is lossless then $\mathbf{G}^\dagger(\omega)\mathbf{G}(\omega)=\mathbf{1}_n$ or that $\mathbf{G}(\omega)$ is unitary
    \item The dispersion relations of Theorem (\ref{tCausality}) hold with $m=1$
\end{itemize}
Furthermore, $\mathbf{G}(\omega)$ is the boundary value of the Laplace transform $\mathbf{G}(k)$ which satisfies the following for all $s>0$
\begin{itemize}
    \item $\mathbf{G}(k)$ is holomorphic
    \item $\mathbf{Q}(k)=\mathbf{1}_n-\mathbf{G}^\dagger(k)\mathbf{G}(k)$ is non-negative definite
    \item $\mathbf{G}^*(k)=\mathbf{G}(k^*)$
\end{itemize}
\end{theorem}

For transfer functions in the immittance form such as $\mathbf{z,y}$, we have a set of similar results as well. These results are only given in terms of $\mathbf{z}$ and its transforms, but it is understood that exactly the same results hold for the admittance as well:

\begin{theorem}\label{tCPImm}
If $\mathbf{z}$ is a matrix of distributional transfer functions corresponding to a linear, passive, and causal system in the immittance form, with its elementwise distributional transforms given by $\mathbf{Z}(\omega),\mathbf{Z}(k)$, then all of the following are true for $s>0$
\begin{itemize}
    \item $\mathbf{Z}(k)$ is holomorphic
    \item $\mathbf{Z}^\dagger(k)+\mathbf{Z}(k)$ is non-negative definite
    \item $\mathbf{Z}^*(k)=\mathbf{Z}(k^*)$
\end{itemize}
Furthermore
\begin{itemize}
    \item $\mathbf{Z}(k)$ has the boundary value $\mathbf{Z}(\omega)$ as $s\rightarrow 0$
    \item $\mathbf{Z}^\dagger(\omega)+\mathbf{Z}(\omega)$ is non-negative definite
    \item The dispersion relations of Theorem (\ref{tCausality}) hold with $m=2$
\end{itemize}
\end{theorem}

The immittance results also have a connection to the so called Herglotz or Nevanlinna functions \citep{herglotz1911uber}. If $\mathbf{Z}$ was a scalar, $Z$, then it would satisfy holomorphicity as well as $\Im (iZ)>0$ in the region $s>0$. These are precisely the conditions for $iZ$ to be a Herglotz function and the following theorem applies to Herglotz functions.

\begin{theorem}\label{tHerglotz}
Necessary and sufficient condition for $R(k)$ to be a Herglotz function is that there exists a bounded non-decreasing real function $\beta(\omega')$ such that
\begin{eqnarray}
R(k)=Ak+C+\int_{-\infty}^{\infty}\frac{1+\omega'k}{\omega'-k}d\beta(\omega');\quad s>0
\end{eqnarray}
where $A,C$ are real constants and $A\geq 0$. Furthermore
\begin{eqnarray}
R(k)/k\rightarrow A\quad \mathrm{as}\quad |k|\rightarrow\infty
\end{eqnarray}
\end{theorem}

{A tensorial equivalent of the above result also exists and was given by Youla \citep{youla1958representation} (See Lemma 4 in Beltrami's paper \citep{beltrami1967linear})}.

\section{Dispersion Relations}\label{sec:dispersion}

At this point, it is clear that the most general form of the dispersion relation, as demanded by causality, is given by Theorem (\ref{tCausality}). It is, therefore, of value to consider them in more detail. For now, we reproduce the relation with the subtraction constants included:
\begin{eqnarray}
\label{eHilbertGenDisSub}
\mathbf{W}(\omega)=-\frac{\omega^n}{\pi i}\left[\frac{\mathbf{W}(\omega)}{\omega^n}*\mathcal{P}\left(\frac{1}{\omega}\right)\right]+\mathbf{P}_{n-1}(\omega)
\end{eqnarray}
where $\mathbf{P}_{n-1}(\omega)$ is a matrix of appropriate size, with each of its elements being a polynomial of degree $\leq n-1$ in $\omega$. In the above, it is clear that the subtraction constants are not needed if one makes use of analytic continuation, in which case $\mathbf{P}_{n-1}(\omega)=\mathbf{0}$ \citep{guttinger1966generalized}. While the subtraction constants are mathematically not needed in the dispersion relations, historically they have been utilized in various areas of physics where they are used as additional parameters which need to be determined through experiments. If $\mathbf{W}(\omega)$ is a matrix of ordinary functions, then Eq. (\ref{eHilbertGenDisSub}) is:
\begin{eqnarray}
\label{eHilbertGenDisSubExpand}
\mathbf{W}(\omega)=\frac{\omega^n}{\pi i}\mathcal{P}\int_{-\infty}^{\infty}\left[\mathbf{W}(\omega')-\mathbf{W}(0)-...-\frac{\omega^{'n-2}}{(n-2)!}\mathbf{W}^{(n-2)}(0)\right]\frac{d\omega'}{\omega^{'n}(\omega'-\omega)}+\mathbf{P}_{n-1}(\omega),
\end{eqnarray}
which is equivalent to the point function with subtractions result of Eq. (\ref{ePlemeljSubtractionN}). The extra terms appear from the process of subtracting out the divergent part of the integral (see Appendix A in \citep{nussenzveig1972causality}). From this point on, we will write the dispersion relations without the subtraction constants unless we are talking about specific areas where they have been used, while keeping in mind that the constants themselves are often simply seen as additional fitting parameters which need to be determined. In any case, as far as practical applications of the dispersion relations are concerned, there does not appear to be an overwhelming consensus on whether the constants should be used or not, with them being added or dropped rather arbitrarily. Furthermore, when talking formally about the dispersion relation, we will also suppress all the terms inside the square brackets in Eq. (\ref{eHilbertGenDisSubExpand}) except for $\mathbf{W}(\omega')$. Again, in practical applications of the dispersion relations, these terms are sometimes ignored on arguments (often physically sound) that they are zero at the chosen frequency. 

Irrespective of $n$, we can immediately use the dispersion relation to connect the real and imaginary parts of the transfer function matrix to arrive at the relations below which apply element-wise:
\begin{eqnarray}
\label{eKramersKronigDistMat}
\Re \mathbf{W}(\omega)=\frac{\omega^n}{\pi}\mathcal{P}\int_{-\infty}^{\infty}\frac{\Im \mathbf{W}({\omega^{'}})}{\omega^{'n}}\frac{d\omega'}{\omega'-\omega};\quad
\Im \mathbf{W}(\omega)=-\frac{\omega^n}{\pi}\mathcal{P}\int_{-\infty}^{\infty}\frac{\Re \mathbf{W}({\omega^{'}})}{\omega^{'n}}\frac{d\omega'}{\omega'-\omega}
\end{eqnarray}
If $\mathbf{w}(t)$ is real, then the integrals can be restricted to positive frequencies, since in that case $\mathbf{W}(-\omega)=\mathbf{W}^*(\omega)$:
\begin{eqnarray}
\label{eKramersKronigDistMatPositiveF}
\nonumber \Re \mathbf{W}(\omega)=\frac{2\omega^n}{\pi}\mathcal{P}\int_{0}^{\infty}\frac{\omega^{'}\Im \mathbf{W}({\omega^{'}})}{\omega^{'n}}\frac{d\omega'}{\omega^{'2}-\omega^2};\quad \Im \mathbf{W}(\omega)=-\frac{2\omega^n}{\pi}\mathcal{P}\int_{0}^{\infty}\frac{\omega\Re \mathbf{W}({\omega^{'}})}{\omega^{'n}}\frac{d\omega'}{\omega^{'2}-\omega^2};\quad n\; \mathrm{even}\\
\Re \mathbf{W}(\omega)=\frac{2\omega^n}{\pi}\mathcal{P}\int_{0}^{\infty}\frac{\omega\Im \mathbf{W}({\omega^{'}})}{\omega^{'n}}\frac{d\omega'}{\omega^{'2}-\omega^2};\quad \Im \mathbf{W}(\omega)=-\frac{2\omega^n}{\pi}\mathcal{P}\int_{0}^{\infty}\frac{\omega^{'}\Re \mathbf{W}({\omega^{'}})}{\omega^{'n}}\frac{d\omega'}{\omega^{'2}-\omega^2};\quad n\; \mathrm{odd}
\end{eqnarray}

As has been mentioned earlier, $n$ is connected to our knowledge of how the quantities of interest behave in the limit $|\omega|\rightarrow \infty$. If the quantities go to 0, then $n=0$ will be sufficient in the above relations to ensure the convergence of the integrals. However, if the quantities are merely bounded by a constant, then $n=1$, at the minimum, will be required. If any $n=m$ is sufficient in a given application, then dispersion relations for all $n>m$ are also applicable. Writing dispersion relations of order higher than needed has the benefit of better convergence of the dispersion integrals \citep{nussenzveig1972causality}. 

\subsection{Examples of Dispersion Relations}\label{subsec:examples}

Dispersion relations have been applied to numerous areas of physics. The first step in identifying the quantities on which dispersion relations apply is to identify those linear, time-translation invariant cause-effect relationships which must necessarily be causal from a physical perspective. The second step is to identify how these causal transfer functions behave in the limit $|\omega|\rightarrow \infty$.

As mentioned earlier, in electrical networks, admittance and impedance are causal because they relate physical quantities through such relations. In electromagnetism, electrical permittivity $\epsilon$ relates electrical displacement to electrical field and must be causal (analytic in the upper half). For similar reasons, magnetic permeability $\mu$ must also be analytic. Since electric and magnetic susceptibilities, $\chi_e,\chi_m$, are simply related to $\epsilon,\mu$, they are also analytic. Furthermore, since the index of refraction $n'=\sqrt{\epsilon\mu}$, its \emph{square} is also analytic in a straightforward manner. What is more difficult to prove is the analyticity of $n'$, since it is not a transfer function between any two physical quantities, and the square root of an analytic function is not necessarily analytic. The index of refraction appears in the expression of a monochromatic plane wave propagating through the medium $\exp[i\omega((n'/c_0) x-t)]=\exp[i(\kappa x-\omega t)]$, where $c_0$ is a constant and $\kappa(\omega)$ is the complex wavenumber. Therefore, the index of refraction is related to the complex wavenumber in a direct manner $n'\propto \kappa(\omega)/\omega$. One can construct a model problem \citep{nussenzveig1972causality} where this wave strikes one side of a slab of a finite thickness and emerges on the other side, and frame an expression of causality based on this -- the wave cannot emerge on the other side before sufficient time has passed after the arrival of the incident wave (relativistic causality). This expression of causality assumes that information cannot travel faster than some constant speed $c_0$, and it is sufficient to show that $n'$ is causal as well. {Skaar \citep{skaar2006fresnel} has shown that in the case of materials with gain, causality does not necessarily imply that $n'$ is analytic in the upper half. In general in such materials, the lack of analyticity does not necessarily imply a loss of causality but instead could imply a lack of stability \citep{milton2020further}.} For the slab problem, part of the incident wave is reflected back with a complex amplitude $r(\omega)=(n'-1)/(n'+1)$, which can also be shown to be an analytic quantity \citep{bode1940relations,jahoda1957fundamental}.

In problems of acoustics, density $\rho(\omega)$ and the bulk modulus $B(\omega)$ must be causal since they relate physical quantities. The quantities $n',\kappa(\omega)/\omega$ are directly related to the quantity $\sqrt{\rho/B}$ and are, therefore, not causal through straightforward arguments. Furthermore, within the framework of acoustics, there is no limiting velocity as there exists in electromagnetism and, therefore, the causality of $n'$ must be proven through other means. However, once $n'$ is proven to be causal, then its square ($\rho/B$) is automatically causal. Below we discuss some examples of dispersion relations which appear in various causal systems in physics.

\subsubsection{Dispersion Relations Applied to Material Properties}

The earliest examples of the application of Kramers-Kronig relations are in the field of wave propagation \citep{Kronig1926,kramers1927diffusion}. The basic ideas which underpin their application in various domains of wave propagation are similar \citep{weaver1981dispersion}. Consider a 1-D plane wave given by the usual form $A\exp[i(\kappa x-\omega t)]$, where $\kappa(\omega)$ is the complex frequency dependent wavenumber of the wave. The refractive index of the medium is generally related to the wavenumber using a relation of the type $n'\propto \kappa(\omega)/\omega$. Although the proof is not straightforward, it can be shown that both $\kappa(\omega)$ and $\kappa(\omega)/\omega$ are analytic in the upper half \citep{nussenzveig1972causality}. Therefore, there is a requirement that $n'$ is also analytic in the upper half. If we can surmise the behavior of either $n'$ or $\kappa(\omega)$ in the high frequency limit, then it should be straightforward to determine the exact form of the dispersion relation which applies to these quantities. What we can say about these limiting quantities depends upon the kind of the wave under consideration.

For electromagnetic waves, $n'(\omega)$ is proportionally related to $\sqrt{\epsilon(\omega)}$ (assuming that the magnetic permeability, $\mu$, is equal to unity). $\epsilon(\omega)$ is in turn related to the susceptibility of the medium $\chi(t)$, which relates the physical quantities electric polarization and electric field through a convolution relation. $\chi(t)$ is automatically causal from physical considerations and, therefore, $n'(\omega),\kappa(\omega)/\omega$ are causal as well. Furthermore, since the dielectric is underlined by a vacuum and the high frequency behavior of a wave approaches that of vacuum propagation \citep{nussenzveig1972causality,weaver1981dispersion}, physics dictates that in the high frequency limit, $\epsilon(\omega)$ tends to $1$ (since $\chi(\omega)$ goes down as $1/\omega^2$ and $\epsilon=1+4\pi\chi$). Since $\epsilon(\omega)$ tends to 1 in the high frequency limit, so does $n'(\omega)$. For electromagnetic wave propagation, $n'(\omega)=n_r+i(c_0\beta/2\omega)$, where the real part of $n'$, $n_r$, is called the real refractive index, and the factor $\beta$ is called the extinction coefficient which governs the attenuation of the medium. $c_0$ is the speed of light which is a constant. In the high frequency limit, since $n'(\omega)\rightarrow 1$, we have $n'(\omega)-1$ tending to zero. Therefore, dispersion relations with 0 subtractions apply to $n'(\omega)-1$ \citep{nussenzveig1972causality}:
\begin{eqnarray}
\label{eKramersKronigDistRefractiveIndex}
n_r(\omega)-1=\frac{c_0}{2\pi }\mathcal{P}\int_{-\infty}^\infty \frac{\beta(\omega')}{\omega'(\omega'-\omega)}d\omega',
\end{eqnarray}
thus linking the real refractive index to the extinction coefficient. Mandelstam \citep{mandelstam1962dispersion} has discussed dispersion relations for $n'(\omega)-1$ under different asymptotic behaviors in the high frequency limit. In those cases, he arrived at dispersion relations which essentially correspond to the subtraction cases (or $n>0$) discussed above. {However, we note an important property of electromagnetic refractive index here: it must always converge to unity in order to satisfy relativistic causality. Therefore, Mandelstram's concerns in this specific context are academic.} Dispersion relations have had a profound impact in the determination of dielectric properties of various materials. One way to do so is to measure the frequency dependent reflection amplitude, $|r(\omega)|$, in an experiment where a thin film is irradiated with electromagnetic waves in a normal direction. The complex reflectance, $r(\omega)=|r|e^{i\phi}$, is related to $n'$, and if one could determine the phase of $r(\omega)$, $\phi$, then one would be able to determine both $n_r(\omega)$ and $\beta(\omega)$, thus determining $\epsilon(\omega)$. It turns out that while it may be difficult to directly measure $\phi$, dispersion relations apply to the real and imaginary parts of $\mathrm{ln}|r|+i\phi$ \citep{bode1945network,jahoda1957fundamental} - a fact that has been used to determine the optical properties in a range of materials \citep{philipp1959optical,taft1961optical,ehrenreich1962optical,kowalski1990optical,miller1993use,steeman1997numerical,van2016better} (see \citep{lucarini2005kramers,peiponen1998dispersion} for more detailed discussions). It must be noted that one must be careful about treating the singularities of $\mathrm{ln}|r|$ as far as these amplitude-phase dispersion relations are concerned \citep{burge1974applicationa,plieth1975kramers,kop1997kramers}. In general, different measurements are required in experiments conducted in different electromagnetic spectra, thus giving rise to a need for slight variations in the dispersion relations to be applied, and also for the development of sophisticated techniques for data assimilation \citep{philipp1964optical,shiles1980self}. Furthermore, dispersion relations are also available for off-normal reflectance measurements \citep{berreman1967kramers}.

Shortly after the application in electromagnetism, dispersion relations were derived for acoustic waves by Ginzberg \citep{ginzberg1955} (see also \citep{mangulis1964kramers}), however, there are some salient differences from the electromagnetic case. The central questions are the same: is $\kappa(\omega)/\omega$ analytic in the upper half, and what is its behaviour in the high frequency limit? In analogy with susceptibility in electromagnetism, a causal function $s(t)$ can be defined which connects acoustic pressure to particle velocity through a convolution relation. This function is causal from physical arguments. Balance of linear momentum dictates that its Fourier transform, $S(\omega)$, is connected to $\kappa(\omega)/\omega$ through a relation $\kappa(\omega)/\omega=-S(\omega)$, thus establishing the analyticity of $\kappa(\omega)/\omega$. As far as the high frequency behavior of $\kappa(\omega)/\omega$ is concerned for acoustics though, there is no easy analogue of the electromagnetic result. Ginzberg \citep{ginzberg1955} essentially assumed that $\kappa(\omega)/\omega$ exists as $|\omega|\rightarrow\infty$, and that it approaches some limiting value independent of $\mathrm{arg}\omega$, which allowed him to derive the dispersion relations. This issue is also present for wave propagation in solids where the analyticity of $\kappa(\omega)/\omega$ can be proved using similar arguments as for acoustic waves. The high frequency behavior of $\kappa(\omega)/\omega$ depends upon the high frequency behavior of the Fourier transform of the stiffness tensor $\mathbf{C}(\omega)$ \citep{weaver1981dispersion}. However, it does not make sense to talk about the high frequency behavior of $\mathbf{C}(\omega)$ because in the high frequency limit, the continuum approximation breaks down. One solution to this conundrum is to follow Ginzberg \citep{ginzberg1955} and assume that there exists some high frequency limit to $\kappa(\omega)/\omega$. In fact, this is precisely what is done by Futterman \citep{futterman1962dispersive} in his application of dispersion relation to seismic wave propagation (see also \citep{lamb1962attenuation,strick1967determination,azimi1968impulse,randall1976attenuative,liu1976velocity} for further discussions on dispersion in seismic waves and connections to Kramers-Kronig relationships). He derived dispersion relations for the complex refraction index defined as $n'(\omega)=\kappa(\omega)/(\omega/c)$, where $\kappa(\omega)$ is the complex wavenumber as discussed above, and $c$ is the nondispersive speed of seismic wave propagation in the low frequency limit. He argues that it is difficult to envision that the structure of the Earth would resonate to a disturbance at infinite frequency. This allows him to say that the imaginary part of $n'$, which is proportional to attenuation, must be 0 in that limit and the real part must equal some constant $n_r(\infty)$. He then considers the quantity $\Delta n'=n'-n_r(\infty)$ which, by its construction, goes to 0 in the high frequency limit, and derives the dispersion relations with no subtractions for $\Delta n'$. He writes the dispersion relations for two frequencies:
\begin{eqnarray}
\label{eKramersKronigDistSeismic1}
\Re[n'(\omega)-n_r(\infty)]=\frac{1}{\pi}\mathcal{P}\int_{-\infty}^\infty \frac{\Im n'(\omega)}{\omega'-\omega}d\omega';\quad \Re[n'(0)-n_r(\infty)]=\frac{1}{\pi}\mathcal{P}\int_{-\infty}^\infty \frac{\Im n'(\omega)}{\omega'}d\omega'
\end{eqnarray}
which, after subtraction, eliminates the unknown value of the refractive index at infinity:
\begin{eqnarray}
\label{eKramersKronigDistSeismic2}
\Re[n'(\omega)-n'(0)]=\frac{\omega}{\pi}\mathcal{P}\int_{-\infty}^\infty \frac{\Im n'(\omega)}{\omega(\omega'-\omega)}d\omega'
\end{eqnarray}
Since the low frequency behavior of seismic wave propagation is experimentally known, he could further argue that $n'(0)=1$, thus simplifying the dispersion relations even further.

For acoustic wave propagation, the derivation of the correct form of the dispersion relations is often based upon assuming a functional form for attenuation \citep{hamilton1970sound,horton1974dispersion,horton1981comment}. Consider $\kappa(\omega)=\omega/c(\omega)+i\alpha(\omega)$, where $c(\omega)$ is the phase velocity of the wave, and $\alpha(\omega)$ is the attenuation constant. For media in which the attenuation satisfies a frequency power law, $\alpha(\omega)=\alpha_0|\omega|^y$, the number of subtractions to be applied depends upon the power coefficient $y$ \citep{waters1999kramers,waters2000application,waters2003differential,waters2005causality}. For $0<y<1$, dispersion relations with 1 subtraction are applicable:
\begin{eqnarray}
\label{eKramersKronigDistAcoustics}
\frac{1}{c(\omega)}=\frac{2}{\pi}\alpha_0\mathcal{P}\int_0^\infty \frac{\omega^{'y}}{\omega^{'2}-\omega^2}d\omega^{'}
\end{eqnarray}
For higher values of $y$, dispersion relations with higher number of subtractions are applicable and have been published \citep{waters2000application}. It is notable that frequency power law attenuation was thought to be incompatible with Kramers-Kronig relationships till rather recently \citep{szabo1994time,szabo1995causal,he1998simulation}. It is indeed not compatible with dispersion relations with no subtractions, but it is fully compatible with dispersion relations with higher number of subtractions, as well as with dispersion relations based upon distribution theory. In the end, the precise minimum number of subtractions needed for waves in solids and liquids may be a moot point since one could always take more than the absolute minimum number of required subtractions and write a valid dispersion relation. As far as the broad field of waves is concerned, very similar treatments based on causality have been proposed for wave propagation in, among many other applications, sediments and sea water \citep{horton1974dispersion,horton1981comment}, poro-elastic media \citep{beltzer1983kramers,beltzer1983wave,brauner1985kramers}, suspensions \citep{mobley1998ultrasonic}, biological material \citep{waters2005kramers,droin1998velocity,anderson2008interference}, and visco-elastic solids \citep{pritz2005unbounded,parot2007applications,rouleau2013application}. Especially notable are early works by O'Donnell \citep{odonnell1981kramers} and Booij et al. \citep{booij1982generalization} who specialized the Kramers Kronig analysis to dissipation and dispersion in liquids and solids.

\subsubsection{Dispersion Relations Applied to Scattering}\label{sec:scattering}

\begin{figure}[htp]
\centering
\includegraphics[scale=1.0]{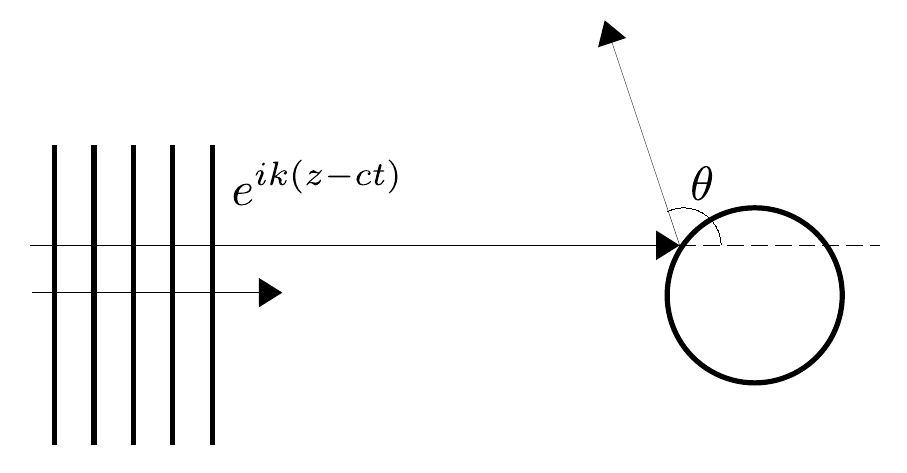}
\caption{Scattering from a spherically symmetrical obstacle}\label{fScatter}
\end{figure}

Scattering matrix was first introduced by Heisenberg \citep{heisenberg1943observable} (although it was discussed even earlier, in passing, by Wheeler \citep{wheeler1937molecular}. See \citep{cushing1986importance} for a fascinating historical criticism of Heisenberg's original program) as a device to describe scattering processes without necessarily referring to the scattering object. Shortly afterwards, Kronig surmised that causality considerations must apply to the elements of the scattering matrix as well \citep{kronig1946supplementary}. The connection lies in the fact that macroscale phenomena such as dispersion and absorption of waves depend ultimately upon the microscale phenomenon of scattering of waves from obstacles. If causality applies to the macroscale properties, then one should be able to formulate causality principles for wave scattering as well whose essential physics is encapsulated in the scattering matrix.

Dispersion relations were first applied to the scattering matrix by van Kampen in a set of two papers \citep{VanKampen1953a,VanKampen1953b} which concerned electromagnetic fields as well as non relativistic particles. To fix ideas as succinctly as possible, however, we summarize the scalar wave case described by Nussenzveig \citep{nussenzveig1972causality}. Acoustic wave scattering reduces to the scalar wave case through the use of velocity potential, and electromagnetic scattering can also be reduced to scalar wave scattering through the use of Debye potentials. Consider scalar wave scattering corresponding to a scalar field $\psi(\mathbf{r},t)$ satisfying the usual scalar wave equation ($\Delta\psi=(1/c^2)\partial^2\psi/\partial t^2$) in a region $r>a$. A plane wave $e^{-i\kappa(z-ct)}$ (which is a solution to the homogeneous scalar wave equation) is scattered by a spherical obstacle of radius $r=a$ centered at the origin (Fig. \ref{fScatter}). We are primarily interested in the response of the system in the far field limit, $r\rightarrow\infty$, which can be shown to be of the following form (suppressing $t$):
\begin{eqnarray}
\label{eScatteringAmplitude}
\psi(\mathbf{r})=e^{-i\kappa z}+f(\kappa,\theta)\frac{e^{-i\kappa r}}{r};\quad r\rightarrow\infty
\end{eqnarray}
Here, the quantity $f(\kappa,\theta)$ is called the scattering amplitude in the $\theta$ direction. It is related to the scattered flux in an infinitesimal solid angle $d\Omega$ through
\begin{eqnarray}
\label{eDifferentialCS}
\frac{d\sigma}{d\Omega}=|f(\kappa,\theta)|^2,
\end{eqnarray}
and, therefore, related to the experimentally measurable total scattering cross-section through
\begin{eqnarray}
\sigma_t(\kappa)=\int|f(\kappa,\theta)|^2d\Omega.
\end{eqnarray}
We have the following important relation called the optical theorem \citep{feenberg1932scattering} which relates the forward scattering to the total scattering cross-section:
\begin{eqnarray}
\label{eOpticalTheorem}
\sigma_t(\kappa)=\frac{4\pi}{\kappa}\Im f(\kappa,0).
\end{eqnarray}
The above expression for the optical theorem is for a lossless scatterer. In the presence of loss, scattering cross-section $\sigma_t$, absorption cross-section $\sigma_a$, and extinction cross-section $\sigma_e=\sigma_t+\sigma_a$ are defined, and the optical theorem holds for $\sigma_e$ instead of $\sigma_t$ \citep{newton2013scattering}. In any case, both $e^{-i\kappa z}$ and $f(\theta)$ can be expanded in Eq. (\ref{eScatteringAmplitude}) in terms of partial waves to give:
\begin{eqnarray}
\label{eAsymptoticExpansion}
\psi(\mathbf{r})\approx \sum_{l=0}^\infty \frac{i^{l+1}}{\kappa}\left[\frac{e^{-i\kappa r}}{r}-(-1)^lS_l\frac{e^{i\kappa r}}{r}\right];\quad r\rightarrow\infty.
\end{eqnarray}
Here, $S_l=1+2if_l$ are the scattering functions, and
\begin{eqnarray}
f_l(\kappa)=\frac{\kappa}{2}\int_0^\pi f(\kappa,\theta)P_l(\cos\theta)\sin\theta d\theta.
\end{eqnarray}
Here, $P_l$ are Legendre Polynomials. In the asymptotic expression (\ref{eAsymptoticExpansion}), index $l$ represents the degree of the partial wave, with $l=0$ corresponding to the spherically symmetric scalar wave solution. Therefore, for a spherically symmetric obstacle, the total field $\psi$, which is generated in response to an incident plane wave, is made up of a set of incoming waves (proportional to $e^{i\kappa r}/r$), and corresponding outgoing waves (proportional to $e^{-i\kappa r}/r$) in the asymptotic limit. These asymptotic limits are, in fact, $r\rightarrow\infty$ limits of spherical Hankel functions which appear in the solution of the problem for all $r>a$. For now, it suffices to note that the total solution is comprised of a discrete set of incoming and outgoing waves which are indexed by $l$. $S_l(\kappa)$ represent scattering functions which connect outgoing wave amplitudes to their respective incoming wave amplitudes. The reality of the fields enforces specific symmetries on $S_l(\kappa)$, energy conservation enforces unitarity properties on $S_l(\kappa)$, and causality ensures that the real and imaginary parts of $S_l(\kappa)$ are not independent of each other. 

To see this more clearly, we can consider the $l=0$ solution independently and note that if there was an $l=0$ incoming wave with a sharp front striking the obstacle at $t=t_0$, then the outgoing cannot be created before $t=t_0$. This is especially clear for the $l=0$ case because the $l=0$ solution maintains its spherically symmetric shape for all $r$, whereas all other modes (for $l>0$) diverge from the spherically symmetric shape as they get closer to $r=a$. Consider, for instance, a combination of $l=0$ incoming and outgoing modes:
\begin{eqnarray}
\label{eSWave}
\psi_0(\kappa,r,t)=\left[A_0(\kappa)(e^{-i\kappa r}/r)+B_0(\kappa)(e^{i\kappa r}/r)\right]e^{-i\kappa ct},
\end{eqnarray}
with the first term defining the incoming wave, and the second term defining the outgoing wave. $S_0(\kappa)=-B_0(\kappa)/A_0(\kappa)$ defines the scattering function for the spherically symmetric scalar wave. We can now build up an incident wave packet with a sharp front:
\begin{eqnarray}
\label{eSWaveIncident}
\psi_{in}(r,t)=\frac{1}{r}\int_{-\infty}^{\infty}A_0(\kappa)e^{-i\kappa c(t+r/c)}d\kappa
\end{eqnarray}
The wave front at the surface of the scatterer is:
\begin{eqnarray}
\label{eSWaveIncidentrequalsa}
\psi_{in}(a,t)=\frac{1}{a}\int_{-\infty}^{\infty}A_0(\kappa)e^{-i\kappa c(t-t_0)}d\kappa,
\end{eqnarray}
where $t_0=-a/c$. We insist that the incident wave reaches $r=a$ at $t=t_0$ such that $\psi_{in}(a,t)$ is 0 for $t<t_0$. The incident wave interacts with the obstacle and gives rise to a scattered wave through the scattering function $S_0$:
\begin{eqnarray}
\label{eSWaveScattered}
\psi_{sc}(r,t)=-\frac{1}{r}\int_{-\infty}^{\infty}S_0(\kappa)A_0(\kappa)e^{i\kappa(r-ct)}d\kappa
\end{eqnarray}
Since both $\psi_{in},\psi_{sc}$ have to be real, we get the symmetry relation $S_0(-\kappa)=S_0^*(\kappa)$. Furthermore, since the incident energy must equal the scattered energy, we get the unitarity condition $|S_0(\kappa)|^2=S_0(\kappa)S_0^*(\kappa)=1$. The unitarity condition means that $S_0(\kappa)$ is simply a phase factor which can be defined by $S_0(\kappa)=e^{2i\eta_0(\kappa)}$ where $\eta_0(\kappa)$ is the phase shift which encapsulates the entire scattering effect of the obstacle on the $l=0$ mode. Evaluated at $r=a$, we have:
\begin{eqnarray}
\label{eSWaveScatteredrequalsa}
\psi_{sc}(a,t)=\frac{1}{a}\int_{-\infty}^{\infty}S_0(\kappa)A_0(\kappa)e^{i\kappa c(a/c-t)}d\kappa=\frac{1}{a}\int_{-\infty}^{\infty}e^{2i\kappa a}S_0(\kappa)A_0(\kappa)e^{-i\kappa c(t-t_0)}d\kappa.
\end{eqnarray}
Causality lies in insisting that if $\psi_{in}(a,t)=0$ for $t<t_0$, then $\psi_{sc}(a,t)$ (Eq. \ref{eSWaveScatteredrequalsa}) must also be 0 for $t<t_0$ (scattered wave must not appear at the obstacle before the incident wave has reached it.) This essentially means that dispersion relations with 1 subtraction can be derived for $e^{2i\kappa a}S_0(\kappa)$ (subtraction at $\kappa=0$):
\begin{eqnarray}
\label{eScatteredDispersion1}
\bar{S}_0(\kappa)=S_0(0)+\frac{\kappa}{\pi i}\mathcal{P}\int_{-\infty}^{\infty}\frac{\bar{S}_0(\kappa')-{S}_0(0)}{\kappa'(\kappa'-\kappa)}d\kappa',
\end{eqnarray}
where $\bar{S}_0(\kappa)=e^{2i\kappa a}S_0(\kappa)$ with generally $S_0(0)=1$. It's possible for derive similar dispersion relations for $S_l(\kappa)$, which is the scattering function for the $l^\mathrm{th}$ partial wave. Waves with $l>0$ do not have a spherical front for small values of $r$ which makes defining a causality condition at $r=a$ troublesome. However, these waves do have spherical fronts at large $r$ and a causality condition can be defined there. The final relations are very similar with symmetry and unitarity of $S_l(\kappa)$ and dispersion relations with one subtraction applicable to $e^{2i\kappa a}S_l(\kappa)$.

Yet another set of dispersion relations can be derived for the scattering cross-section. These dispersion relations are more fundamental than those derived for the scattering functions since the latter follows if the former holds but not the other way around. There is a causality condition associated with $f(\kappa,\theta)$ since the time of arrival of the scattered wave along any angle $\theta$ is related to the time at which the incident wave undergoes a specular reflection at the obstacle (Fig. \ref{fScatter}). If we define:
\begin{eqnarray}
\bar{f}(\kappa,\theta)=e^{2i\kappa \sin(\theta/2)}f(\kappa,\theta),
\end{eqnarray}
then it can be shown that dispersion relations with two subtractions apply to $\bar{f}(\kappa,\theta)$ (after some simplifications):
\begin{eqnarray}
\label{eDispersionF}
\Re\bar{f}(\kappa,\theta)=f(0,\theta)+\frac{2\kappa^2}{\pi}\mathcal{P}\int_{0}^\infty \frac{\Im\bar{f}(\kappa,\theta)}{\kappa^{'}(\kappa^{'2}-\kappa^2)}d\kappa^{'}.
\end{eqnarray}
Given the optical theorem (Eq. \ref{eOpticalTheorem}), this means that the dispersion relation (\ref{eDispersionF}) takes a particularly simple form for $\theta=0$ (forward scattering) \citep{karplus1955applications}:
\begin{eqnarray}
\label{eDispersionF0}
\Re f(\kappa,0)=f(0,0)+\frac{\kappa^2}{2\pi^2}\mathcal{P}\int_{0}^\infty \frac{\sigma_t(\kappa^{'})}{\kappa^{'2}-\kappa^2}d\kappa^{'}
\end{eqnarray}
Furthermore, it can be shown that $\sigma_t(0)=4\pi[f(0,0)]^2$, so that the right hand side of (\ref{eDispersionF0}) can be expressed purely in terms of the scattering cross-section, which is an experimentally measurable quantity. Once $\Re f(\kappa,0)$ has been calculated, $\Im f(\kappa,0)$ can also be calculated since dispersion relations apply to the real and imaginary parts of $f(\kappa,0)$ \citep{rohrlich1952forward}, which allows for the calculation of the differential cross-section in the forward direction (Eq. \ref{eDifferentialCS}). Causality also implies that there exist lower bounds on $d\eta_l/d\kappa$ which are the derivatives of the scattering phase shifts \citep{wigner1955lower}. These relationships have had a profound impact in the design of cloaking devices based upon metamaterial principles (to be discussed later).

van Kampen \citep{VanKampen1953a} reduced the electromagnetic wave propagation problem to the scalar wave problem using the Debye potentials so that the above analysis applies to EM waves. Acoustic wave scattering also follows closely the scalar wave case and was clarified by Hackman \citep{hackman1993acoustic}, although relativistic causality considerations do not apply there as there is no theoretical maximum velocity. The quantum mechanical setting is similar to the scalar wave equation, but the causality treatment there is complicated by the fact that a sharp quantum mechanical wave-front is impossible to create (as the integral analogous to Eq. \ref{eSWaveIncident} in the quantum mechanical setting only runs over the positive values of the parameter). We simply note here that alternative causality statements for nonrelativistic quantum mechanical scattering exist, beginning with the pioneering works from Schutzer and Tiomno as well as van Kampen, which allow us to arrive at dispersion relations \citep{schutzer1951connection,VanKampen1953b,wong1957dispersion,khuri1957analyticity,khuri1958dispersion}. Dispersion relations for relativistic elementary particle physics are also available beginning from the pioneering works of Gell-Mann and others \citep{gell1954use,goldberger1955causality,goldberger1955application,karplus1955applications}.

\subsection{Sum rules, DAR, and nearly local approximations}\label{subsec:approx}

Once the relevant dispersion relations have been identified for a particular problem, there are several further consideration that can be applied to them. An early mathematical result relevant to dispersion relations, which is of far reaching generality is the superconvergence theorem \citep{de1966sum}, which says that if
\begin{eqnarray}
\label{eSuperConvergence}
g(y)=\mathcal{P}\int_0^\infty\frac{f(x)}{y-x}dx
\end{eqnarray}
where $f(x)$ is a continuously differentiable function (beyond some large value $x_0$) which vanishes at infinity faster than $x^{-1}$, then we have:
\begin{eqnarray}
g(y)\approx\frac{1}{y}\int_0^\infty f(x)dx;\quad y\rightarrow\infty.
\end{eqnarray}
The superconvergence theorem applied to the dispersion relations immediately results in a variety of so called sum-rules. As an example, if a medium behaves as a free electron gas in the high frequency limit, then $n'(\omega)-1\approx-\frac{1}{2}\omega^2_p/\omega^2$, where $\omega_p$ is the plasma frequency, and in this case Eq. (\ref{eKramersKronigDistRefractiveIndex}) applies. Writing $n'=n_r+in_i$, where $n_i=c\beta/2\omega$, the dispersion relations with no subtractions are:
\begin{eqnarray}
n_r(\omega)-1=\frac{2}{\pi}\mathcal{P}\int_0^{\infty}\frac{\omega^{'}n_i(\omega^{'})}{\omega^{'2}-\omega^2}d\omega^{'};\quad n_i(\omega)=-\frac{2\omega}{\pi}\mathcal{P}\int_0^{\infty}\frac{n_r(\omega^{'})-1}{\omega^{'2}-\omega^2}d\omega^{'}
\end{eqnarray}
Superconvergence applied to both integrals immediately results in the following useful sum rules \citep{altarelli1972superconvergence}:
\begin{eqnarray}
\int_0^\infty\omega n_i(\omega)d\omega=\frac{1}{4}\pi\omega_p^2;\quad \int_0^\infty [n_r(\omega)-1]d\omega=0
\end{eqnarray}
In their original papers, Altarelli et al. \citep{altarelli1972superconvergence,altarelli1974superconvergence} have formulated sum rules for $n_r(\omega)-1$, the real part of the dielectric tensor, and for gyrotropic media. Sum rules have been formulated for optical constants \citep{kubo1972kramers,smith1976superconvergence,villani1973superconvergent,bassani1991dispersion}, scattering \citep{drell1966exact,maximon1974sum,king1976sum} reflectance \citep{king1979dispersion,ellis1975sum}, strong interactions \citep{de1966sum,gilman1968strong}, nuclear reactions \citep{teichmann1952sum}, and negative refractive index materials \citep{peiponen2004kramers} among other applications. While sum rules contain less information than the dispersion relation itself, they often relate simple integrals over experimentally measurable quantities and have historically led to surprising bounds on such quantities \citep{purcell1969absorption}. 

All the dispersion relations above are integral relationships and, as such, while the real (imaginary) part of causal transfer functions may be calculated from the imaginary (real) part, one needs to know the imaginary (real) part for the entire semi-infinite frequency spectrum in order to do so. In general, however, the real and/or imaginary parts are only known over some finite frequency range which complicates the application of the dispersion relations. The way that this is generally handled is by assuming some form of the integrand outside the measured frequency range, which then enables the calculation of the infinite integrals. For example, in acoustics this is directly done by assuming that the attenuation is related to frequency using a power law type of relationship \citep{horton1974dispersion} $\alpha(\omega)=\alpha_0|\omega|^y$. For optics, the electrical permittivity may expressed as a sum over classical Lorentz oscillators, essentially employing a curve fit in dispersion analysis \citep{spitzer1961infrared,verleur1968determination}:
\begin{eqnarray}
\label{eLorentz}
\epsilon(\omega)=\epsilon_\infty+\sum_{j=1}^N\frac{s_j}{\omega_j^2-\omega^2-i\Gamma_j\omega}
\end{eqnarray}

Another approach is to truncate the infinite integral to a finite range, $D=[\omega_{min},\omega_{max}]$, within which appropriate data is available. For example, it may be that for a causal function $f(\omega)=f_r(\omega)+if_i(\omega)$, the measurement of $f_i$ is only available for $\omega\in D$, and one may still wish to apply dispersion relations over a truncated integral over $D$ in order to calculate $f_r$. This process introduces uncertainty and errors in the results of the dispersion analysis \citep{mobley2000kramers,mobley2003finite}, but the nature of the Hilbert transform is such that it is dominated by the region around $\omega'\approx\omega$, where the kernel is singular. Although it is possible for features outside of $D$ to have arbitrary influence in the interior, it requires a large amount of energy to do so \citep{dienstfrey2001analytic}. Therefore, under certain conditions the truncated integral can give a good approximation to $f_r$. As an example, for reflectance dispersion analysis, Bowlden and Wilmshurst \citep{bowlden1963evaluation} have shown that the error associated with truncation is small if the frequency $\omega$ is far from $\omega_{min},\omega_{max}$. Furthermore, the error is small over the entire range $D$ if the reflectance does not vary sharply outside of the range of interest. This conclusion is another restatement of the observation by Dienstfrey and Greengard \citep{dienstfrey2001analytic}. A similar result exists in acoustics where O'Donnell et al. \citep{odonnell1981kramers} have shown that the truncated integral is a good approximation to the infinite integral if the attenuation factor does not vary sharply outside of the truncation range. Milton et al. \citep{milton1997finite} have provided dispersion bounds for finite frequency dispersion analysis which depend upon the measurement of $f_i$ over $D$, in addition to the measurement of $f_r$ at some discrete $\omega^{(i)}$ points. If both $f_r,f_i$ are known over $D$ (or at least over some overlapping region), then it is possible to recapture $f(\omega)$ over the entire real line \citep{hulthen1982kramers}. However, in spite of the formal existence of such an analytic continuation \citep{aizenberg1993carleman}, it turns out to be a very ill-posed problem \citep{dienstfrey2001analytic}. However, once the analytic continuation is determined, it can be used to calculate partial sum rules \citep{kuzmenko2007model} from limited data.

Another approach towards dealing with finite spectrum data is to convert the integral dispersion relationships to a derivative form (derivative analytic relationships, DAR). Consider a dispersion relation with no subtractions:
\begin{eqnarray}
\Im \mathbf{W}(\omega)=-\frac{1}{\pi}\mathcal{P}\int_{-\infty}^\infty \frac{\Re\mathbf{W}}{\omega'-\omega}d\omega'
\end{eqnarray}
With the substitution $x^{'}=\mathrm{ln}(\omega^{'}/\omega)$, the integral can be transformed into \citep{odonnell1981kramers}:
\begin{eqnarray}
\Im \mathbf{W}(\omega)=-\frac{1}{\pi}\int_{-\infty}^\infty \frac{d\Re\mathbf{W}(x)}{dx}\mathrm{ln\;coth}\left(\frac{|x|}{2}\right)dx
\end{eqnarray}
$\mathrm{ln\;coth}\left(\frac{|x|}{2}\right)$ is sharply peaked at $x=0$, and, thus, the magnitude of the integral is dominated by the values of the integrand around $x=0$. It is, therefore, possible to expand $d\Re\mathbf{W}(x)/dx$ around $x=0$ and evaluate the integral resulting in:
\begin{eqnarray}
\label{eDerivativeApproximation}
\Im \mathbf{W}(\omega)=-\frac{\pi}{2}\frac{d\Re\mathbf{W}(x)}{dx}\Biggr|_{x=0}-\frac{\pi}{24}\frac{d^3\Re\mathbf{W}(x)}{dx^3}\Biggr|_{x=0}...
\end{eqnarray}
which may be further simplified by considering only the first term in the series in the absence of sharp variations in $\Re\mathbf{W}(x)$. The above treatment is due to O'Donnell et al. in the context of acoustics, however, the original derivative relations come from Bronzan et al. \citep{bronzan1974obtaining}, who derived these relationships for what is essentially the $n=1$ case. This $n=1$ case corresponds to the once subtracted dispersion relations, and Bronzan et al. \citep{bronzan1974obtaining} presented their discussion within the context of relating the real and imaginary parts of the scattering amplitude in high energy collision physics applications \citep{eden1967high}. In these applications, the once subtracted relations are used to improve the convergence of the integrals due to the behavior of the quantities as $\omega\rightarrow\infty$ \citep{block1985high}. The original derivative relations have since undergone much scrutiny and commentary in the high energy physics literature \citep{fischer1976derivative,fischer1978high,kolavr1984validity,fischer1987differential,avila2004critical,ferreira2008representation}. The relationships presented below are based upon a generalization of the original derivative analyticity relations, which were given by Waters et al. \citep{waters2003differential} whose main interest was acoustics (see also \citep{menon1999differential}):
\begin{eqnarray}
\label{eKramersKronigDistMatPositiveFDAR}
\nonumber \Re \mathbf{W}(\omega)=\omega^{n-1}\tan\left[\frac{\pi}{2}\frac{d}{d\mathrm{ln}\omega}\right]\frac{\Im \mathbf{W}(\omega)}{\omega^{n-1}};\quad \Im \mathbf{W}(\omega)=-\omega^{n}\tan\left[\frac{\pi}{2}\frac{d}{d\mathrm{ln}\omega}\right]\frac{\Re \mathbf{W}(\omega)}{\omega^{n}};\quad n\; \mathrm{even}\\
\Re \mathbf{W}(\omega)=\omega^{n}\tan\left[\frac{\pi}{2}\frac{d}{d\mathrm{ln}\omega}\right]\frac{\Im \mathbf{W}(\omega)}{\omega^{n}};\quad \Im \mathbf{W}(\omega)=-\omega^{n-1}\tan\left[\frac{\pi}{2}\frac{d}{d\mathrm{ln}\omega}\right]\frac{\Re \mathbf{W}(\omega)}{\omega^{n-1}};\quad n\; \mathrm{odd}
\end{eqnarray}
{The tangent functions in the above encapsulate the infinite order derivatives present in \ref{eDerivativeApproximation}, which themselves are the result of a Taylor series expansion, as mentioned earlier.} It is clear that the complexity of evaluating the infinite integrals in Eqs. (\ref{eKramersKronigDistMatPositiveF}) has been exchanged for the complexity of evaluating the infinite derivative orders in Eqs. (\ref{eKramersKronigDistMatPositiveFDAR}). Furthermore, the integral and the derivative relationships are not exactly analogous because the derivative relationships require that the tangent series be convergent -- an assumption that has faced considerable criticism \citep{eichmann1974can,heidrich1975derivative,bujak1976there,kolavr1984validity}. The derivative analyticity relations simplify further if one makes the assumption that the quantities of interest do not vary too quickly with $\omega$. This is true when the $\omega$ under consideration is away from resonance phenomenon (Sukhatme et al. \citep{sukhatme1975extensions} have discussed simplifications to the relations when this assumption is relaxed, however, not without criticisms primarily related to convergence of the tangent series in the first place \citep{bujak1976there}). In such cases, it is sufficient to retain only the first derivative term and the relations reduce to:
\begin{eqnarray}
\label{eKramersKronigDistMatPositiveFDARLocal}
\nonumber \Re \mathbf{W}(\omega)\approx\omega^{n-1}\frac{\pi}{2}\frac{d}{d\mathrm{ln}\omega}\frac{\Im \mathbf{W}(\omega)}{\omega^{n-1}};\quad \Im \mathbf{W}(\omega)\approx-\omega^{n}\frac{\pi}{2}\frac{d}{d\mathrm{ln}\omega}\frac{\Re \mathbf{W}(\omega)}{\omega^{n}};\quad n\; \mathrm{even}\\
\Re \mathbf{W}(\omega)\approx\omega^{n}\frac{\pi}{2}\frac{d}{d\mathrm{ln}\omega}\frac{\Im \mathbf{W}(\omega)}{\omega^{n}};\quad \Im \mathbf{W}(\omega)\approx-\omega^{n-1}\frac{\pi}{2}\frac{d}{d\mathrm{ln}\omega}\frac{\Re \mathbf{W}(\omega)}{\omega^{n-1}};\quad n\; \mathrm{odd}
\end{eqnarray}
which is generally called a nearly local approximation to the dispersion relations. The nearly local relation due to O'Donnell et al. (Eq. \ref{eDerivativeApproximation}) discussed above, when constrained to the first term in the context of acoustics (relating attenuation $\alpha(\omega)$ with the phase velocity $c(\omega)$), results in:
\begin{eqnarray}
\alpha(\omega)=\frac{\pi\omega^2}{2c^2(\omega)}\frac{dc(\omega)}{d\omega}
\end{eqnarray}
which turns out to be a special case of Eq. (\ref{eKramersKronigDistMatPositiveFDARLocal}) for $n=1$ \citep{waters2003differential}. {Note that Eqs. (\ref{eKramersKronigDistMatPositiveFDARLocal}) are merely approximations and should be used with care. For instance they suggest that there exist local constraints on the real and imaginary parts of causal $\mathbf{W}$ separately, which is not always the case.}

\section{Causality, Passivity, and Metamaterials}\label{sec:metamaterials}

Metamaterials are artificially designed composite materials which can exhibit properties that can not be found in nature. The field is very broad and it is not our purpose to review it. We refer to other reviews for the same \citep{craster2012acoustic,vendik2013metamaterials,hussein2014dynamics,turpin2014reconfigurable,Srivastava2015ElasticReview,cummer2016controlling,wang2016optical,ren2018auxetic,yu2018mechanical}. Here, we focus on three broad strains in metamaterials research which are pertinent to the current topics of causality and passivity.

\begin{itemize}
    \item Design and creation of composite materials which exhibit unusual material properties. These are permittivity and permeability ($\epsilon(\omega),\mu(\omega)$) in electromagnetism, density and bulk modulus ($\rho(\omega),B(\omega)$) in acoustics, and density and stiffness tensor ($\rho(\omega),\mathbf{C}(\omega)$) in elastodynamics. All properties can be possibly tensorial and one of the original goals of the field was to achieve negative wave refraction which is facilitated by negative effective properties \citep{pendry2000negative,liu2000locally}.
    \item Determination of the above material properties through scattering simulations and experiments \citep{srivastava2014limit,amirkhizi2017homogenization}.
    \item Design of scattering devices, such as cloaks, which are targeted towards the manipulation of the scattering cross-section $\sigma_t$. \citep{greenleaf2003anisotropic,greenleaf2003nonuniqueness,leonhardt2006notes,leonhardt2006optical,pendry2006controlling,cummer2006full,milton2006cloaking,norris2008acoustic,norris2011elastic}. 
\end{itemize}

\subsection{Metamaterial Properties and the Lorentz Oscillator Model}
The original goal of metamaterials research was to create optical materials which would exhibit simultaneously negative $\epsilon,\mu$. Such a material would possess a negative refractive index $n'=-\sqrt{\epsilon\mu}$ from arguments of causality \citep{veselago1968electrodynamics}. If, in addition, we could have $\epsilon=\mu=-1$, then light would pass through such a material without reflection and this idea could be used to beat the diffraction limit and create super-lenses \citep{pendry2000negative}. We note here that Pendry's original idea of a superlens, while extremely popular in metamaterials research, nevertheless suffers from serious deficiencies, some of which are related to causality \citep{mcphedran2019review}. In any case, it is notable that $\epsilon,\mu=-1$ (more generally $\epsilon,\mu<0$) is admissible within a causal framework. Smith et al. \citep{smith2000negative} alluded to this idea based upon original arguments by Landau and Lifshitz \citep{landau2013electrodynamics}, however, it is also apparent from the fact that susceptibilities for general dispersive, non-absorbing systems consist of a sum of causal Lorentz contributions of the form given in Eq. (\ref{eLorentzOscillatorGamma0}) \citep{tip1998linear,tip2004linear,gralak2010macroscopic}:
\begin{eqnarray}
\label{eDrude}
\epsilon(\omega)=\mu(\omega)=1-\frac{\omega_p^2}{\omega^2-\omega_0^2}-\frac{i\pi\omega_p^2}{2\omega_0}\left[\delta(\omega+\omega_0)-\delta(\omega-\omega_0)\right]
\end{eqnarray}
which both achieve a value of -1 at $\omega^2=\omega_0^2+\omega_p^2/2$. If both $\epsilon,\mu$ are negative then the notion that one must choose the negative root for $n'$ is based upon the requirement that the work done by an electromagnetic source must be a positive quantity \citep{smith2000negative} (see also \citep{akyurtlu2010relationship} for an argument from causality). The same requirement of positive work done ensures that the impedance of the medium, $z=\sqrt{\mu/\epsilon}$, must be positive even when both $\epsilon,\mu<0$. In general, Smith et al. \citep{smith2000negative} argued that $\epsilon,\mu,n',z$ are all causal transforms even in negative index materials, and that all causally consistent materials which exhibit negative $\epsilon,\mu$ must also exhibit frequency dispersion -- a result also noted in the original Veselago paper \citep{veselago1968electrodynamics}. This can also be seen by employing the low-loss approximation inequality originally due to Landau and Lifshitz \citep{landau2013electrodynamics} {(note that the following inequality holds only if $\epsilon(\omega)$ does not have a pole at $\omega=0$ which is not the case with metals. See also \citep{abdelrahman2020broadband})}:
\begin{eqnarray}
\label{eLLInequality2}
\frac{d\epsilon(\omega)}{d\omega}\Biggr|_{\omega_0}\geq\frac{2(1-\epsilon(\omega_0))}{\omega_0};\;\omega_0>0;\;\Im\epsilon(\omega_{1}\leq \omega_0\leq \omega_{2})= 0
\end{eqnarray}
If $\epsilon(\omega_0)<0$ then $d\epsilon(\omega)/d\omega$ is bounded from below by a positive quantity, thus ensuring dispersion. It must also be noted that this dispersion result only applies to passive media. In active media, it is possible to achieve $\epsilon=\mu=-1$ over a finite bandwidth \citep{skaar2006bounds,lind2009perfect}. 

As far as practical realization of negative index materials is concerned, the basic idea is to create resonances in the spectrum of $\epsilon,\mu$ through appropriate design of resonators. If the resonances can be made to coincide in the frequency domain, then one arrives at a negative index material \citep{smith2000composite,pendry1996extremely,pendry1999magnetism}. As long as $\epsilon(\omega),\mu(\omega)$ emerge from equations similar to Eq. (\ref{eLorentz}), they are guaranteed to be causal. In general, however, causality requirements appear to be violated in widely used homogenization models for metamaterials \citep{alu2011causality}.

\begin{figure}[htp]
\centering
\includegraphics[scale=0.6]{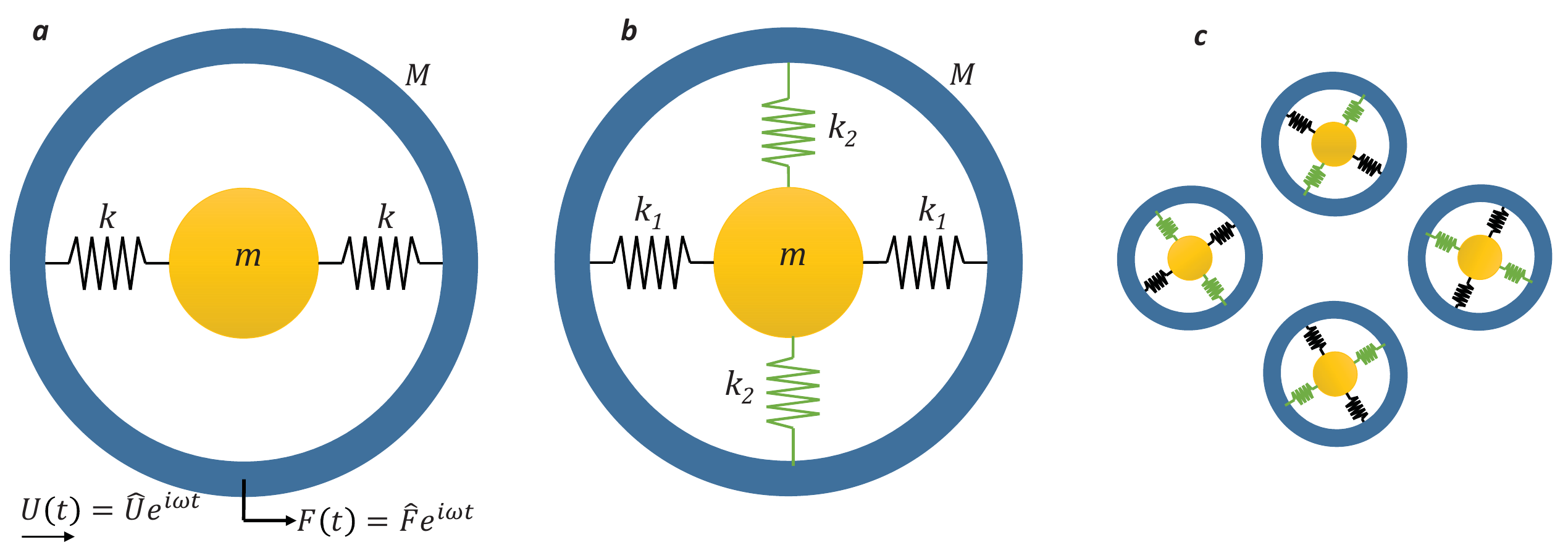}
\caption{A mass in mass system which gives rise to dispersive and possibly anisotropic effective mass and density \citep{milton2007modifications,huang2009negative,Srivastava2015ElasticReview}}\label{fND}
\end{figure}

Interestingly enough, parallel developments in acoustics were not based on similarly rigorous grounds as those in electromagnetic metamaterials. The early papers in the area of acoustic metamaterials, while highly influential no doubt, are largely experimental \citep{liu2000locally,sheng2003locally}. They were trying to show that something as fundamental as density could be dispersive and the idea that it could become negative does not seem to be as much of a focus \citep{mei2006effective}. It appears that the fact that naturally occurring optical materials exhibit $\epsilon$ which is of the form given by Eq. (\ref{eLorentz}), whereas no naturally occurring material has density of the same form, has resulted in a lag in parallel developments in the area of acoustic metamaterials. This lag was addressed when Milton and Willis \citep{milton2007modifications} showed that mechanical resonances give rise to effective density which is in the Lorentz form (see also \citep{huang2009negative}):
\begin{eqnarray}
\rho(\omega)=\rho_0+\sum_{j=1}^N \frac{s_j}{\omega_j^2-\omega^2-i\Gamma_j\omega}
\end{eqnarray}
While Milton and Willis \citep{milton2007modifications} did not show how to achieve negative modulus, they did elaborate upon the dispersive (and anisotropic) nature of both density and modulus through homogenization theory (see \citep{Srivastava2015ElasticReview} for a relevant review). Fang et al. \citep{fang2006ultrasonic} showed that effective modulus can also be made to assume the Lorentz form using Helmholtz resonators, and shortly afterwards these ideas were combined to produce double negative acoustic materials \citep{ding2007metamaterial,cheng2008one,lee2010composite} {(Note that discussions on negative and anisotropic density and stiffness have a longer history in homogenization literature \citep{auriault1985,auriault1994,schoenberg1983properties}.)} It is of note to consider that while rigorous arguments based upon causality were made in support of choosing the appropriate sign of $n',z$ for electromagnetic metamaterials, no such arguments have been made for acoustic metamaterials.

\subsection{Are Metamaterial Properties through the Retrieval Method Causally Consistent?}

How does one assign effective metamaterial properties to heterogeneous structures? One way to do so is through coherent averaging principles (homogenization) applied to the full-field solutions of the heterogeneous structures with appropriate boundary conditions \citep{willis1997dynamics,craster2010high}. Another method is to extract these properties from an appropriate scattering simulation or experiment. From the perspective of causality and passivity, it is the latter which is of interest to us. In its simplest form, the setup involves measuring the complex valued reflection coefficient, $r$, and transmission coefficient, $t$, as a normally incident wave is scattered by a thin plate of a known thickness, $d$. The plate itself is made up of some unit cells of the heterogeneous structure whose effective properties need to be determined. If the plate was made up of a homogeneous material with index of refraction, $n'$, and impedance, $z$, then the relations between $r,t$ and $n',z$ are given by the Fresnel-Airy formulas. By inverting the Fresnel-Airy formulas we get:
\begin{eqnarray}
\label{eFresnelAiry}
\nonumber \kappa n'=\pm\frac{1}{d} \mathrm{cos}^{-1}\left(\frac{1-r^2+t^2}{2t}\right)+\frac{2\pi m}{d}\\
z=\pm\sqrt{\frac{(1+r)^2-t^2}{(1-r)^2-t^2}}
\end{eqnarray}
where $\kappa$ is the wavenumber of the incident wave and $m$ is an integer function of frequency. Therefore, measurement of $r,t$ allows for the numerical calculation of $n',z$, which then may be used to assign effective material properties to the entire plate through simple relations. This retrieval technique has been used in numerous papers in both electromagnetism \citep{smith2002determination} and acoustics \citep{fokin2007method}, however, it is not our concern here to review all those papers. What is important for our purpose is to note that there is some amount of discretion involved in the retrieval methods. It lies in the appropriate choice of the parameter $m$ as well as the signs. Generally, this choice is employed so as to make the calculated effective properties passive, which requires that their imaginary parts be positive for positive frequencies. However, it does nothing to make the calculated effective properties causally consistent. In other words, there is nothing in the retrieval method which ensures that the calculated real and imaginary parts of the effective properties satisfy the Kramers-Kronig dispersion relations. In fact, Simovski \citep{simovski2009material} has argued that a majority of the papers which use retrieval methods in metamaterials research end up calculating effective properties which violate causality. The reason for this is fundamental and relates to the presence of a layer of evanescent waves at the interface between the metamaterial plate and the surrounding medium \citep{Srivastava2017EvanescentApproach} -- a phenomenon which is necessarily ignored in retrieval methods which try to assign homogeneous effective properties to the entire finite region of the scatterer. In doing so, they are essentially trying to apply periodic averaging to a phenomenon which is not periodic. Due to this fundamental issue, it may be asserted that metamaterial properties calculated from retrieval methods, in general, will violate causality. 

\subsection{Must Negative Index Materials be Dissipative?}

An interesting question is whether causality demands that negative index materials must be inherently lossy, and whether this loss can be compensated for by gain while maintaining negative properties. Stockman \citep{stockman2007criterion}, by applying dispersion relations to $n^{'2}$, arrived at the following equality which must be satisfied in order to simultaneously have negative refraction and zero loss at a frequency $\omega$:
\begin{eqnarray}
\label{eStockman2007}
\frac{c^2}{v_pv_g}=1+\frac{2}{\pi}\int_0^\infty\frac{\Im n^{'2}(s)}{(s^2-\omega^2)^2}s^3ds
\end{eqnarray}
where $v_p,v_g$ are phase and group velocities respectively. For media which exhibits negative refraction, Stockman took $v_pv_g<0$, which allowed him to derive an inequality which must be satisfied by the real and imaginary parts of $\epsilon,\mu$. He showed that this inequality ensures that significantly reducing, by any means passive or active, the losses associated with negative refraction resonances will annihilate the negative refraction itself. Stockman \citep{stockman2007stockman} has asserted that while it is possible to have zero or low losses at some isolated frequencies, it is impossible to remove losses in the entire region of negative refraction without destroying the negative refraction itself. This result has received criticism in literature \citep{mackay2007comment,mackay2009dispersion}. Stockman's results depend upon his assumption that $\Im n^{'2}$ and its derivative are zero at the observation frequency as the conditions for vanishing loss. This requirement could be relaxed. Kinsler \citep{kinsler2008causality} has presented another more general causality based inequality, involving $\Im n^{'2}$ and its first two derivatives, which must be satisfied by all negative index media. As a more emphatic comment, however, on the question of whether all negative index media need to be lossy -- the answer is no. This is due to the observation that the simple Lorentz oscillator model of Eq. (\ref{eLorentzOscillator}) with vanishing loss, $\gamma\rightarrow 0^+$, has a transfer function which satisfies the dispersion relations (Eq. \ref{eLorentzOscillatorGamma0}) \citep{poon2009kramers}, and allows for real negative values of the transfer function. So, in theory, one could fashion both $\epsilon,\mu$ from causal Lorentz models with vanishing loss and arrive at a media which exhibits finite frequency bandwidths with negative index behavior and no loss, all the while being causally consistent \citep{gralak2010macroscopic,tip2004linear,milton2020further}. In other words, dissipation need not accompany dispersion (even in the negative property regime and over finite bandwidths) from a causality perspective \citep{nistad2008causality}.

\subsection{Constraints on Metamaterials from Passivity and Causality}

What is the effect of passivity requirements on metamaterial properties? To answer this, we can refer back to the results of section \ref{sec:Passive}. For diagonal (or scalar) metamaterial properties, passivity demands that the imaginary parts of the diagonal values of these be non-negative for all positive values of the frequency $\omega$ \citep{schwinger1998classical,tip1998linear,liu2013causality}. For more general cases, Srivastava \citep{Srivastava2015CausalityElastodynamics} has considered a system characterized by a tensorial and distributional transfer function $\mathbf{L}$ and satisfying the very general input-output relation $\mathbf{x}=\mathbf{L}*\mathbf{f}$ such that its absorbed energy is given by:
\begin{equation}
\Re\;\mathcal{E}(t)=\Re\int_{-\infty}^{t}\mathrm{d}s\;\mathbf{w}^\dagger(s)\dot{\mathbf{v}}(s)=\Re\int_{-\infty}^{t}\mathrm{d}s\; \mathbf{w}^\dagger(s)\int_{-\infty}^{\infty}\mathrm{d}v\; \dot{\mathbf{L}}(v)\mathbf{u}(s-v),
\end{equation}
with the passivity statement that $\Re\mathcal{E}(t)\geq 0\;\forall t$. Such a passivity statement appears naturally in electromagnetism, acoustics, and elastodynamics. Srivastava has shown that this passivity statement imposes the following positive semi-definiteness requirements on the Fourier and Laplace transform of $\mathbf{L}$ and its time derivatives (see also \citep{konig1958lineare}):
\begin{equation}\label{positivedefSG1}
\mathbf{y}^\dagger\tilde{\dot{\mathbf{L}}}^h\mathbf{y}\geq 0;\quad\mathbf{y}^\dagger\hat{\dot{\mathbf{L}}}^h\mathbf{y}\geq 0;\quad \mathbf{y}^\dagger\omega\tilde{{\mathbf{L}}}^{nh}\mathbf{y}\geq 0,
\end{equation}
where \emph{tilde} denotes the Fourier transform and \emph{hat} denotes the Laplace transform (right half convention considered in the original paper). The superscripts $h,nh$ denote the hermitian and non-hermitian (with the factor $i$ removed) parts respectively. This result is sufficiently general to apply to electromagnetism, acoustics, and elastodynamics as special cases, in addition to Willis property tensors \citep{willis2009exact,willis2011effective,Nemat-Nasser2011OverallComposites,srivastava2012overall,willis2012construction}, which can be thought of as the superset of all these cases {(see \citep{pernas2020fundamental} for an application to the generalized Willis tensor in systems including piezoelectric effects)}. For the passivity statement $\Re\mathcal{E}(t)\geq 0\;\forall t$, it follows that the results of Theorem \ref{tCPImm} apply to $\tilde{\dot{\mathbf{L}}},\hat{\dot{\mathbf{L}}}$. Furthermore, if $\mathbf{L}$ was a scalar $L$ then the relations in (\ref{positivedefSG1}), in combination with discussions from section \ref{sec:Passive} would mean that $\omega\tilde{L}(\omega)$ would be a Herglotz function (Theorem \ref{tHerglotz}). As shown earlier, passivity is fully consistent with causality so one obviously expects that a material satisfying passivity will have material properties which  satisfy dispersion relations. We have also seen earlier that if the relevant properties are tensor-valued, then the dispersion relations apply elementwise (Eq. \ref{eHilbertGenDisSub}). Muhlestein et al. \citep{muhlestein2016reciprocity} have discussed these dispersion relations in the context of homogenized Willis kinds of relationships and they arrive at dispersion relations which are essentially the analogues of  Eq. (\ref{eHilbertGenDisSub}) with zero subtractions, as well as some of the associated derivative relationships discussed earlier.

Gustafsson and Sjoberg \citep{gustafsson2010sum} have used the fact that $\omega L$ is a Herglotz function to create sum rules and bounds on $\epsilon(\omega)$. Before we talk about these, we refer back to some early inequalities derived by Laundau and Lifshitz \citep{landau2013electrodynamics} which must also be satisfied by metamaterial properties. They argued that with $\epsilon(\infty)=1$ and with the assumption that $\Im\epsilon(\omega_0)=0;\forall\omega_0\in[\omega_1,\omega_2]$, the derivative of $\epsilon(\omega)$ can be bounded from below using several inequalities, the sharpest of which is presented in Eq. (\ref{eLLInequality2}). In a more general form, the inequality is:
\begin{eqnarray}
\label{eLLInequality2Modified}
\frac{d\epsilon(\omega)}{d\omega}\Biggr|_{\omega_0}\geq\frac{2(\epsilon(\infty)-\epsilon(\omega_0))}{\omega_0}
\end{eqnarray}
It's possible to integrate the above to get to the following inequality:
\begin{eqnarray}
\mathrm{max}\;|\epsilon(\omega)-\epsilon_m|\geq \frac{\omega_2-\omega_1}{\omega_0}(\epsilon(\infty)-\epsilon_m);\quad \omega,\omega_0\in[\omega_1,\omega_2];\quad \epsilon_m=\epsilon(\omega_0)
\end{eqnarray}
The inequality says that if one wants to achieve a specific value of real $\epsilon_m$ at some $\omega_0\in[\omega_1,\omega_2]$, where $[\omega_1,\omega_2]$ is the region where loss is zero, then the deviations from $\epsilon_m$ in the vicinity of $\omega_0$ are inevitable and, in fact, bounded from below by the above inequality. {It must also be noted that this result is the same as the two-point bound presented by Milton et al. \citep{milton1997finite}.} Gustafsson and Sjoberg \citep{gustafsson2010sum} showed that this Landau Lifshitz derived bound does not apply in the presence of loss ($\Im\epsilon>0$) even if the loss is vanishingly small ($\Im\epsilon\rightarrow 0^+$). For the lossy case, they derived several more general bounds using the fact that $\omega\epsilon(\omega)$ is a Herglotz function. One of those bounds is presented below:
\begin{eqnarray}
\label{egustaffson}
\mathrm{max}\;|\epsilon(\omega)-\epsilon_m|\geq \frac{B/2}{1+B/2}(\epsilon(\infty)-\epsilon_m);\quad \epsilon_m\leq\epsilon(0)
\end{eqnarray}
where $B=(\omega_2-\omega_1)/\omega_0$ {and the maximum is over the interval $[\omega_1,\omega_2]$}. As an explicit example, (assuming $\epsilon(\infty)=1$), the bound above says that if $\epsilon=-1$ is desired at some $\omega_0$, then causality requires that the deviation of $\epsilon$ from $-1$ will be at least $1\%$ in a region $B=1\%$ around $\omega_0$. It may be possible to derive similar bounds for acoustics and elastodynamics but none have been presented yet. {Finite bandwidth bounds have also been presented by Skaar and Seip \citep{skaar2006bounds} and Lind-Johansen et al. \citep{lind2009perfect}. Notably, Lind-Johansen et al. \citep{lind2009perfect} have presented a bound which is similar to (\ref{egustaffson}) but is tight. They showed that the infimum of the $L^\infty-$ norm of $\chi+2$ over the interval $B$ is equal to $2\Delta/(1+\sqrt{1-\Delta^2})$, where $\Delta=(\omega_2^2-\omega_1^2)/(\omega_2^2+\omega_1^2)$. Since the susceptibility $\chi=\epsilon-1$, this infimum is indicative of the divergence of $\epsilon$ from a value of $-1$ over the interval $[\omega_1,\omega_2]$.}

\subsection{Causality Constraints on Scattering from Cloaks}

Another area within metamaterials research in the last two decades where the considerations of causality have been important is in the design of cloaking devices for both electromagnetic waves and acoustic waves. Fleury et al. \citep{fleury2015invisibility} have published an excellent review on the various cloaking mechanisms that have been developed over the last two decades, and they also address performance limitations on cloaks coming from arguments of causality. Not wishing to reiterate the case, here we only discuss the topic succinctly and include some more recent references not included in Fleury et al. \citep{fleury2015invisibility}. The essential goal of a wave cloaking device is to hide some region $\Omega$ from interrogation by a wave. Various strategies could be implemented towards this goal and all of them are geared towards suppressing the scattered field -- the quantity $f(\kappa,\theta)$, or equivalently $f(\omega,\theta)$ in Eq. (\ref{eScatteringAmplitude}) -- produced as the wave impinges on the region $\Omega$. Where does causality enter into this setup? The answer depends upon the kind of strategy being pursued for the design of the cloak.

Miller \citep{miller2006perfect} envisaged an active cloaking strategy where sensors and active sources are distributed on $\partial\Omega$. The sensors measure the incoming wave-field and actuators respond to cancel out the scattering. He concluded that the constraint that electromagnetic information cannot travel faster than the speed of light (relativistic causality) limits the performance of such a device. He showed that in this scheme, perfect cloaking is impossible over finite frequency bandwidths if the actuators respond to only local information (sensors in the vicinity). {It should be noted that good (but imperfect) cloaking may still be achievable in this scheme if the fields vary slowly, in which case it may become possible to predict, with reasonable accuracy, the fields in the near future based upon their past values.} Chen et al. \citep{chen2007extending} considered the causal limitations on cloaks based on coordinate transformation techniques. They showed that causality ensured that perfect cloaking can only be achieved at a single frequency. In fact, material property dispersion required by causality ensures that cloaking performance decreases inversely with both the size of the object being cloaked and the desired bandwidth over which cloaking is desired \citep{hashemi2012diameter}. The cloaking bandwidth could only be increased by tolerating a higher amount of minimum scattering by the cloak --  a result also noticed for plasmonic cloaks by Alu and Engheta \citep{alu2008effects}. In general, passive cloaks are bandwidth limited from causality, a limitation not faced as acutely by active cloaks \citep{chen2019active}.

There is a fundamental issue which places bounds on how much a linear, passive, and causal cloak can really scatter. The issue can be understood by referring to the essential discussions in section (\ref{sec:scattering}). $\sigma_t(\kappa)$ is a measure of the total scattering from the cloak at some frequency $\omega=c\kappa$, and this quantity is directly related to the imaginary part of the forward scattering amplitude $\Im f(\kappa,0)$ through the optical theorem (Eq. \ref{eOpticalTheorem}). We can define the total scattering from the cloak -- a scalar quantity $\Sigma$, also called the \emph{integrated extinction} -- as an appropriate integral of $\sigma_t(\kappa)$ over the entire frequency range (or equivalently $0\leq\kappa\leq\infty$). The optical theorem ensures that $\Sigma$ will be related to the equivalent integral of $\Im f(\kappa,0)$. Now, since dispersion relations (with 2 subtractions) apply to $f(\kappa,0)$ (e.g. Eq. \ref{eDispersionF}), we can express a Hilbert integral on $\Im f(\kappa,0)$ in terms of $\Re f(\kappa,0)$, which may be evaluated at $\kappa=0$. In effect, through a clever choice of $\Sigma$, it is possible to relate a relevant measure of the total scattering from the cloak ($\Sigma$) to the real part of the forward scattering amplitude at 0 frequency (or $\kappa=0$), $\Re f(0,0)$. A passive cloak cannot scatter any less than this value since the actual design does not appear in this chain of logic. Furthermore, since $f(0,0)$ is the forward scattering in the quasi-static regime, it can be expressed directly in terms of the static averages of the material properties of the cloak. The origins of this idea are due to Purcell \citep{purcell1969absorption} who established causality based sum rules on the integrated extinction of electromagnetic waves due to interstellar gases. Gustafsson et al. \citep{gustafsson2007physical} provided the relevant expressions in the context of antennas. These ideas were exploited by Monticone and Al{\`{u}} \citep{monticone2013cloaked} to show that linear, passive, and causal electromagnetic cloaks actually scatter more than uncloaked objects (in the sense of integrated extinction; see also \citep{fleury2014physical}). Note that the optical theorem can also be used to derive upper bounds on scattering from a cloak \citep{liberal2014upper}. {Since passive systems possess a Herglotz representation, Cassier and Milton \citep{cassier2017bounds} have used bounds on Herglotz functions to present fundamental limits on broadband passive quasistatic cloaking. It should be noted that some of the bandwidth limitations due to causality also do not apply to electromagnetic cloaks if the background material in which the cloak is placed is not vacuum \citep{alu2008effects}.}

It is interesting to note that most of these bounds do not apply to acoustic cloaks, since they depend upon the assumption that there is a maximum velocity of information travel -- no such maximum applies for acoustics. This is discussed in detail by Norris \citep{norris2015acoustic,Norris2015}, who has provided sum rules on the integrated extinction even in the absence of causality (see also his more recent paper on integral identities \citep{norris2018integral}). Norris has shown that for acoustic cloaking, $\Sigma$ can vanish for a wide variety of scatterers, including for so called neutral acoustic inclusions.

\section{Conclusions}

In this review, we have discussed the concepts of causality and passivity from a variety of perspectives, and drawn from developments in a range of fields in mathematics, physics and engineering. Our final goal is to understand how the ideas of causality and passivity fit within the modern field of metamaterials, and what are some future potential directions of inquiry. However, the answers to those questions are not easy to ascertain without understanding the vast body of knowledge which already exists on the topics of causality and passivity. This body of knowledge tells us that dispersion relations can be derived for a large class of transfer functions and not just for those which are square integrable. These dispersion relations can then be used to derive sum-rules using powerful theorems such as the superconvergence theorem. The dispersion integrals can sometimes be truncated if one is sure that the quantities involved do not exhibit resonances outside of the truncation integral. Finally, these relations can be expressed as derivative relationships, and under certain conditions these relationships reduce to particularly simple forms which may then be used to judge the causal consistency of a model using only local data. It is well known that there is a close correspondence between causality and passivity. Passivity implies causality but causality does not necessarily imply passivity. Furthermore,  passivity also determines the precise form of the dispersion relation (number of subtraction terms) that is followed by a system's transfer function. 

As far as metamaterials research is concerned, on balance it appears that the development of causally consistent ideas and causality constraints in the field of electromagnetic metamaterials is much more advanced than it is in the parallel fields of acoustic metamaterials and elastic metamaterials. No doubt, part of the reason for this is the existence of a limiting velocity in electromagnetism whereas no such concept exists for acoustics or elastodynamics. Regarding the determination of causally consistent metamaterial properties from reflection/transmission experiments, there again appears to be a difference in the state of development of ideas in the three classical metamaterial areas. As an example, as far as we know, the concept of transition layers has not been developed for acoustic or elastic wave metamaterials. We admit that these transition layers might not be the only way of dealing with the issue. The idea that dispersion relations with higher number of subtractions always apply whenever a lower order subtraction does, has not at all been exploited in metamaterials research. In all, to us it appears that there are fruitful future directions of research at the confluence of causality, passivity, and metamaterials; especially acoustic and elastic wave metamaterials.

\acknowledgments     
 
A.S. acknowledges support from the NSF CAREER grant \#1554033 to the Illinois Institute of
Technology.


\end{document}